\pdfoutput=1
\documentclass[amsmath,amssymb,preprintnumbers,nofootinbib,a4paper,11pt]{article}

\usepackage{graphicx}
\usepackage{epstopdf}
\usepackage{jcappub}
\usepackage{amsmath,amsthm,latexsym,amssymb,amsfonts,epsfig}
\usepackage{fontenc,layout}
\usepackage{wrapfig}
\usepackage{caption}
\usepackage{hyperref}
\usepackage{psfrag}
\usepackage[cp1250]{inputenc}
\usepackage[usenames,dvipsnames]{xcolor}

\newcommand{\be}{\begin{equation}}
\newcommand{\ee}{\end{equation}}
\newcommand{\bea}{\begin{eqnarray}}
\newcommand{\eea}{\end{eqnarray}}
\newcommand{\beq}{\begin{eqnarray}}
\newcommand{\eeq}{\end{eqnarray}}

\numberwithin{equation}{section}

\begin{document}

\title{Gravitational waves from first-order cosmological phase transitions:
lifetime of the sound wave source
}

\begin{flushright}
KCL-PH-TH/2020-04 \\ CERN-TH-2020-016 \\ IFT-UAM/CSIC-20-35
\end{flushright}

\author[1,2,3]{John ~Ellis}
\author[1,4]{, Marek ~Lewicki}
\author[5,6]{, Jos\'e ~Miguel ~No}
\affiliation[1]{Department of Physics, King's College London, Strand, London WC2R 2LS, UK}
\affiliation[2]{Theoretical Physics Department, CERN, Geneva, Switzerland}
\affiliation[3]{National Institute of Chemical Physics \& Biophysics, R\"avala 10, 10143 Tallinn, Estonia}
\affiliation[4]{Faculty of Physics, University of Warsaw, ul.\ Pasteura 5, 02-093 Warsaw, Poland}
\affiliation[5]{Instituto de F\'isica Te\'orica, IFT-UAM/CSIC,
Cantoblanco, 28049 Madrid, Spain}
\affiliation[6]{Departamento de F\'isica Te\'orica, Universidad Aut\'onoma de Madrid, 28049 Madrid, Spain}
\emailAdd{John.Ellis@cern.ch}
\emailAdd{Marek.Lewicki@kcl.ac.uk}
\emailAdd{Josemiguel.no@uam.es}

\abstract{We survey systematically the general parametrisations of particle-physics models for a first-order phase transition in the early universe, including models with polynomial potentials both with and without barriers at zero temperature, and Coleman-Weinberg-like models with potentials that are classically scale-invariant. We distinguish three possibilities for the transition - detonations, deflagrations and hybrids - and consider sound waves and turbulent mechanisms for generating gravitational waves during the transitions in these models, checking in each case the requirement for successful percolation. We argue that in models without a zero-temperature barrier and in scale-invariant models the period during which sound waves generate gravitational waves lasts only for 
a fraction of a Hubble time after a generic first-order cosmological phase transition, whereas it may last longer in some models with a zero-temperature barrier that feature severe supercooling. We illustrate the implications of these results for future gravitational-wave experiments.}

%\notoc

%\begin{document}

\maketitle

%\newpage

%\tableofcontents

\section{Introduction}

Many scenarios for physics beyond the Standard Model (BSM) predict a first-order phase transition in the early universe, around or before the epoch of electroweak symmetry breaking. Studies of such models have long been motivated by the hope of realising electroweak baryogenesis (see, e.g., \cite{Cohen:1993nk,Trodden:1998ym,Morrissey:2012db}). Another motivation has come to the fore recently, namely that a strong first-order phase transition is also a potential source of gravitational waves (GWs) that could be measured by future detectors  (see, e.g.,
~\cite{Kamionkowski:1993fg,Grojean:2006bp,Huber:2008hg,Espinosa:2008kw,Dorsch:2014qpa,Jaeckel:2016jlh,Chala:2016ykx,Chala:2018opy,Artymowski:2016tme,Hashino:2016xoj,Vaskonen:2016yiu,Dorsch:2016nrg,Beniwal:2017eik,Baldes:2017rcu,Marzola:2017jzl,Kang:2017mkl,Chala:2018ari,Vieu:2018zze,Bruggisser:2018mrt,Megias:2018sxv,Croon:2018erz,Alves:2018jsw,Baratella:2018pxi,Angelescu:2018dkk,Croon:2018kqn,Brdar:2018num,Beniwal:2018hyi,Breitbach:2018ddu,Marzo:2018nov,Baldes:2018emh,Prokopec:2018tnq,Fairbairn:2019xog,Helmboldt:2019pan,Dev:2019njv,Jinno:2019jhi,Ellis:2019flb,Jinno:2019bxw,Azatov:2019png,Hindmarsh:2019phv,vonHarling:2019gme,DelleRose:2019pgi}
and also~\cite{Caprini:2015zlo,Caprini:2018mtu,Caprini:2019egz} for recent reviews), allowing a direct probe of the physics of the early Universe before Big-Bang Nucleosynthesis. 
Specifically, a cosmological first-order phase transition during the radiation-dominated era could, depending on the temperature at which it took place, 
%thermally-induced phase transition at the electroweak scale could 
generate a GW signal at frequencies $\sim 10$~Hz where LIGO~\cite{TheLIGOScientific:2014jea} and the future Einstein Telescope (ET)~\cite{Punturo:2010zz} have optimal detection capabilities, and/or the $0.1 - 1$ Hz frequency band where atom interferometers such as MAGIS~\cite{Graham:2017pmn}, AION~\cite{Badurina:2019hst} and AEDGE~\cite{Bertoldi:2019tck} may offer the best hope of detection, and/or the $10^{-3} - 10^{-2}$ Hz frequency band where space-based laser interferometers such as LISA~\cite{Audley:2017drz} are most sensitive (see~\cite{Figueroa:2018xtu} for a recent analysis of the synergies among the various GW detectors). A third motivation for studying BSM scenarios that predict a first-order phase transition in the early universe is the possibility that they might generate a primordial intergalactic magnetic field (see, e.g., \cite{Ellis:2019tjf}) or produce primordial black holes~\cite{Lewicki:2019gmv,Konoplich:1999qq,Moss:1994iq,Hawking:1982ga}.

The above motivations underline the importance of modelling accurately such a first-order cosmological phase transition, and its potential GW signal in particular. This is a non-trivial problem, especially in the interesting case of violent, strong transitions,
since there are several mechanisms for generating GWs, involving the interplay between {scalar} field theory and the hydrodynamics of the primordial plasma.  
Among the sources of GWs from a first-order phase transition are bubble collisions, sound waves and turbulence in the plasma following the transition. The {plasma sound waves that develop after the collisions of 
bubbles~\cite{Hindmarsh:2013xza,Hindmarsh:2015qta,Hindmarsh:2016lnk,Hindmarsh:2017gnf,Hindmarsh:2019phv}} have attracted particular attention (see~\cite{Caprini:2015zlo,Caprini:2019egz} for a review), in view of their distinctive peaked frequency distribution and the fact that they are potentially a long-lasting GW source, which could in general give rise to a sizeable GW signal,
and also in view of the difficulties in estimating reliably the strength of the GW signal generated by turbulence in the plasma. Bubble collisions are an easier source to understand theoretically, since it is not necessary to model the plasma. However, it is not easy to realise a transition strong enough for this source to be important in a thermal background (see, e.g.,~\cite{Ellis:2019oqb}).

A key element in estimating the strength of the GW from sound waves is the {duration of the acoustic period during which sound waves may be emitted.}
%A key element in estimating the strength and shape of the GW signal from sound waves is how long-lasting they are. 
Some calculations have been based on estimates that the period of sound wave emission 
%much longer than the Hubble time
lasts a Hubble time, {after which} Hubble expansion shuts down the GW source (Hubble damping). However, we have found previously~\cite{Ellis:2018mja,Ellis:2019oqb} that in specific models this 
acoustic period lasts significantly less than a Hubble time.
%sound waves are emitted in a relatively short-lived burst, lasting $\lesssim$ a Hubble time. 
%
In this work we discuss systematically the possible duration of the acoustic period
%We show in this paper that this is a general feature 
in a variety of particle-physics scenarios featuring a first-order cosmological phase transition, emphasising that the parameters describing the strength and duration of the transition are not independent, and taking the requirement of successful percolation into account. We consider generic polynomial potential models with and without zero-temperature barriers, as well as Coleman-Weinberg-like potential models that are almost scale-invariant. We find that the acoustic period is necessarily limited in polynomial models without a zero-temperature barrier, and in Coleman-Weinberg-like models. However, it may extend as long as a Hubble time in polynomial models with a zero-temperature barrier, though only in a restricted range of parameter space. We illustrate the implications of these results for the potential GW signals in the LIGO~\cite{TheLIGOScientific:2014jea}, ET~\cite{Punturo:2010zz}, MAGIS~\cite{Graham:2017pmn}, AION~\cite{Badurina:2019hst}, AEDGE~\cite{Bertoldi:2019tck} and LISA~\cite{Audley:2017drz} detectors.

The structure of this paper is as follows.~In Section~\ref{sec:correlation} we review the general ingredients to describe a first order phase transition in the early Universe, including in Section~\ref{sec:hydrodynamics} a discussion of the behaviour of the plasma following the transition, with more details given in Appendix~\ref{sec:Hydro_Appendix}.~In 
Section~\ref{sec:thinwall} we analyse the correlation between the strength and duration of a phase transition in the simple thin-wall approximation.
%, and show that it can be surprisingly reliable. 
We extend the discussion to generic polynomial potentials and Coleman-Weinberg-like potentials in Section~\ref{sec:genral_polynomial}. The implications of our results for the GW signals are illustrated in some specific scenarios in Section~\ref{sec:GWsignals}, and we summarise our conclusions in Section~\ref{sec:conx}.
Relevant aspects of hydrodynamics, details on the completion of the transition
and explicit analytic formulae for parameters of the phase
transitions in the models we study are presented in the Appendices.

%%%%%%%%%%%%%%%%%%%%%%%%%%%%%%%%%%%%%%%%%%%%%%%%%%%%%%%%%%%%%%%%%%%%%%%%%%

\section{Review of phase transition dynamics and parameters}
\label{sec:correlation}

%%%%%%%%%%%%%%%%%%%%%%%%%%%%%%%%%%%%%%%%%%%%%%%%%%%%%%%%%%%%%%%%%%%%%%%%%%%
 
\subsection{Definitions and preliminaries} 
\label{sec:definitions}
 
Our starting point is the decay rate of the false vacuum due to thermal effects~\cite{Linde:1980tt,Linde:1981zj}:
\be 
\label{eq:gamma:2.1}
\Gamma\propto T^4 \exp{-\frac{S_3(T)}{T}}\,,
\ee
where $S_3$ is the action for the tunnelling field solution.\footnote{We refer the reader to~\cite{Ellis:2018mja} for a more detailed discussion.} In the simplest approximation the transition takes place when, on average, one bubble is nucleated in every horizon volume, which defines the transition time (temperature) $t_*$ ($T_*$):
\be \label{eq:nucleation_temperature}
\int_0^{t_*} \frac{\Gamma}{H^3}dt=\int_{T_c}^{T_*}\frac{dT}{T} \frac{\Gamma}{H^4}=1\, ,
\ee
where in the second equality we only integrate from $T_c$ (the temperature at which the two minima are degenerate) since the decay rate vanishes at earlier times. 
The next step in studying the phase transition is to obtain $\Gamma$ as a function of temperature. For a generic scalar field, the action of the tunnelling solution from a false vacuum $\phi_{\rm f}$ to a true vacuum $\phi_{\rm t}$, assuming as usual $O(3)$ symmetry, is given by
\be \label{eq:bounce_action}
S_3=4\pi \int dr r^2 \left( \frac{\dot{\phi}^2}{2} + V(\phi) \right)
\ee
where, in order to find the field profile $\phi=\phi(r)$, one typically has to solve numerically the corresponding equation of motion 
\be
\ddot{\phi}+\frac{2}{r}\dot{\phi}=\frac{d V}{d \phi}
\ee
with the boundary conditions
\be
\dot{\phi}(r=0)=0 \, , \quad \phi(r\rightarrow \infty)=\phi_f \, ,
\ee
whose profile corresponds to a bubble nucleating in the false vacuum background and reaching beyond the barrier at the origin of our coordinates. 

\vspace{1mm}

The inverse duration of the phase transition (in Hubble units), $\beta/H$, is generally found by Taylor-expanding to first order~\footnote{Except for very strong supercooling, for which this method breaks down and further terms in the expansion need to be included in the computation of $\beta/H$ (see~\cite{Ellis:2018mja,Megevand:2016lpr,Jinno:2017ixd,Cutting:2018tjt} for details).} the exponent in Eq.~\eqref{eq:gamma:2.1} around $t_*$
\be
\label{eq:beta_Taylor}
\Gamma \propto  e^{-\frac{S_3(T)}{T}} = e^{\beta(t-t_*) + ...} \, ,
\ee
and differentiating both sides of \eqref{eq:beta_Taylor} with respect to $t$, which yields 
\be
\beta=-\frac{d}{dt}\frac{S_3(T)}{T}=H(T)\,  T\frac{d}{dT}\frac{S_3(T)}{T} \, .
\ee
The strength, $\alpha$, of a first-order phase transition may be defined as the difference in the trace of the energy-momentum tensor $\theta = (e -3 p)/4$ between the symmetric and broken phases, normalised to the radiation background energy density in the unbroken phase 
(see~\cite{Caprini:2019egz} for details), i.e.
\be
\alpha(T) \equiv \frac{\Delta \theta}{\rho_{\mathrm{R}}} = \frac{\Delta V(T) - \frac{T}{4} \frac{\partial \Delta V(T)}{\partial T}}{\rho_{\mathrm{R}}} \, ,
\label{eq:alpha}
\ee
where $\Delta V(T) = V_{\rm f} - V_{\rm t}$, with $V_{\rm f} \equiv V(\phi_{\rm f})$ and $V_{\rm t} \equiv V(\phi_{\rm t})$ the values of the potential in the false and true vacua respectively.

\vspace{2mm}

Before continuing, we note that the two terms in the action (\ref{eq:bounce_action}) are not in fact independent.
Based on simple dimensional analysis one can show that for a bubble profile solution they are in fact identical up to a coefficient~\cite{Coleman:1985rnk}, such that
\be \label{eq:bounce_action_expanded}
S_3=4\pi \int dr \, r^2 \left( \frac{\dot{\phi}^2}{2} + V(\phi) \right)
=\frac{2}{3} \times 4\pi \int dr\, r^2 \, \frac{\dot{\phi}^2}{2}
=-2 \times 4\pi \int dr\, r^2\,  V(\phi) \, ,
\ee 
where we set $V(\phi_{\rm f})=0$, in which case the integral above is finite.
The last equality in~\eqref{eq:bounce_action_expanded} already points to the fact that the value of $\beta/H$, being related to the derivative of the action $S_3$, is correlated with the value of $\alpha$, which is related to the endpoints of the integrand in this last equality.

\subsection{Hydrodynamics}
\label{sec:hydrodynamics}

We now review the main aspects of the plasma hydrodynamics in the presence of expanding bubbles from a cosmological first-order phase transition, as needed for a description of the various GW sources from the transition (see Section~\ref{sec:GWsignals}). As is customary in phase transition studies, we consider a single expanding bubble interacting with the plasma background and follow the derivation in~\cite{Steinhardt:1981ct,Ignatius:1993qn,Espinosa:2010hh} (see 
also~\cite{Hindmarsh:2019phv}), expanding on it when necessary.   

%Before we proceed to treatment of more generic models featuring a cosmological phase transition, let us review  the hydrodynamics necessary for a more complete description of such processes. This allows us calculate the the root-mean-square four-velocity of the plasma that we will then need for a good description of all the sources of gravitational waves. 

\vspace{1mm}

We assume that the thermal plasma can be described as a perfect fluid, with
a four-velocity field
\be 
U_{\mu} = \frac{(1,\,\vec{v})}{\sqrt{1 - \left|\vec{v}\right|^2  }} = (\gamma,\,\gamma \,\vec{v}) \, .
\ee
The energy-momentun tensor then reads $T_{\mu\nu} = w U_{\mu} U_{\nu} - g_{\mu\nu} p$, where $ p $ is the pressure and $w = T \, (\partial p / \partial T)$ is the enthalpy. We can then use energy-momentum conservation, $\partial_{\mu} T^{\mu\nu} = 0$, with appropriate boundary conditions, to derive the behaviour of the plasma.
The conservation of energy and momentum is non-trivial across the bubble wall, due to the change in the pressure $\Delta p = - \Delta V$ between the two phases.
Matching the energy-momentum inside ($-$) and outside ($+$) the bubble gives the conditions satisfied at the bubble wall~\cite{Steinhardt:1981ct}:
%
%(with ``$+$" denoting the symmetric phase and ``$-$" the broken phase)
%
\be \label{eq:matching_conditions} 
\frac{w_{+} \,v^2_{+}}{1-v^2_{+}} + p_{+} = \frac{w_{-}\,v^2_{-}}{1-v^2_{-}} + p_{-} \; \; \;\;\;\; {\rm and} \;\;\; \; \; \; 
\frac{w_{+}\, v_{+}}{1-v^2_{+}} = \frac{w_{-} \, v_{-}}{1-v^2_{-}} \, ,
\ee
where $v = \left|\vec{v}\right|$. 
Assuming a simplified bag-model equation of state,\footnote{This simplified model follows from a relativistic gas approximation, and has been shown also to be a  good approximation in more realistic conditions~\cite{Espinosa:2010hh}.} we obtain the equation describing the difference in plasma velocities on the two sides of the wall: 
\be
v_{+} = \frac{1}{1 + \alpha_{+}} \left[ \left(\frac{v_{-}}{2} + \frac{1}{6 v_{-}}   \right) \pm \sqrt{\left(\frac{v_{-}}{2} + \frac{1}{6 v_{-}}   \right)^2 + 
\alpha_{+}^2 + \frac{2}{3} \alpha_{+} - \frac{1}{3}}\right] \, ,
\label{eq:matching_wall}
\ee
where $\alpha_+$ is normalised to the radiation energy density immediately in front of the bubble wall, and may not coincide with the strength of the transition obtained from Eq.~\eqref{eq:alpha} (we elaborate on the connection between the two in Appendix~\ref{sec:Hydro_Appendix}). 

In order to obtain the plasma velocity and enthalpy profiles away from the phase boundary we again use energy-momentum conservation, which gives
\be
\label{Tmunuplasma_conservation}
\partial_{\mu} T^{\mu\nu} = U^{\nu} \partial_{\mu} (U^{\mu} w) + U^{\mu} w \partial_{\mu} U^{\nu} - \partial^{\nu}p = 0 \, .
\ee
We consider only spherically-symmetric bubble configurations, such that the fluid spatial dependence is given in terms of the radial coordinate $r$ from the centre of the bubble. Since there are no other relevant scales in the problem apart from the time since nucleation, $t$, the velocity and enthalpy/temperature profiles exhibit a self-similar behaviour, characterised by the ratio $\xi = r/t$.
With these conditions, we obtain from~\eqref{Tmunuplasma_conservation}
differential equations for the velocity profile $v(\xi)$ and enthalpy profile 
$w(\xi)$ of the plasma
\be
\label{eq:vfluid}
\frac{2\, v}{\xi} = \frac{1 - \xi\,v}{1- v^2} \left[\frac{1}{c_s^2}\left(\frac{\xi-v}{1-\xi\, v}\right)^2 - 1 \right] \partial_\xi v \, ,
\ee
\begin{equation}
\label{eq:enthalpy}
  \partial_{\xi}  \omega =2 \omega \left(c_s^2+1\right)\frac{ \, v}{\xi}\frac{\xi-v}{\left(\xi-v\right)^2-c_s^2\left(1-\xi  v\right)^2}\, ,
\end{equation}
%
%
% \be
% \gamma^2= \frac{1}{1- v^2} \, , \quad \quad \mu^2(\xi,v)=\left(\frac{\xi-v}{1-\xi\, v}\right)^2
% \ee
where $c_s$ is the speed of sound in the plasma, given by $c_s^2 = 1/3$ in a relativistic fluid.
%This can also be rewritten in an integral form, but it is more efficient simply to solve this differential equation numerically.
%
The appropriate boundary conditions for these equations depend on the 
terminal velocity $v_w$ that the bubble wall reaches.\footnote{In principle this quantity has to be derived from the microphysical description of the interactions between the background scalar field varying across the bubble wall and the thermal plasma in a given particle physics model, see~\cite{Khlebnikov:1992bx,Arnold:1993wc,Moore:1995ua,Moore:1995si,John:2000zq,Megevand:2009gh,Huber:2011aa,Kozaczuk:2015owa,Dorsch:2018pat}. In this work we instead treat $v_w$ as a free parameter, and study the sensitivity to different values.}
As $v_w$ increases, the plasma profile accompanying the expanding bubble may behave in three qualitatively different ways, the so-called {\it deflagration}, {\it hybrid} and {\it detonation} solutions to Eqs.~\eqref{eq:vfluid} and~\eqref{eq:enthalpy}~\cite{Espinosa:2010hh}, whose details we give in Appendix~\ref{sec:Hydro_Appendix}.
From the fluid profiles for $v(\xi)$ and $w(\xi)$ given by these solutions, we can obtain several quantities on which the GW signals from the phase transition depend directly. 
First, the fraction $\kappa$ of energy in the fluid converted into bulk fluid motion (i.e., not gone into reheating the plasma), given by~\cite{Espinosa:2010hh}
\be
\kappa = \frac{3}{\alpha\, \rho_{\mathrm{R}} \,v_w^3} \int w\, \xi^2 \frac{v^2}{1-v^2} d\xi \, .
\label{eq:kappa_eff}
\ee        
%
%where the enthalpy is already normalised to the radiation background of the symmetric phase. 
% We show the resulting efficiency $\kappa$ for several values of $\alpha$ as a function of the wall velocity $v_w$ in Fig.~\ref{fig:kapp_sw}.

% %
% \begin{figure}[h]
% \begin{center}
% \includegraphics[width=0.7\textwidth]{kappaplot.pdf}
% \vspace{-0.2cm}
% \caption{\it Efficiency of energy transfer into bulk fluid motion $\kappa$ for selected values of $\alpha$ as function of the wall velocity $v_w$. The solid black line marks the boundary between deflagrations and hybrids, while the black-dashed line marks the boundary between hybrids and detonations (see Appendix~\ref{sec:Hydro_Appendix} for details).}
% \label{fig:kapp_sw}
% \end{center}

\begin{figure}[h]
\begin{center}
\includegraphics[width=0.49\textwidth]{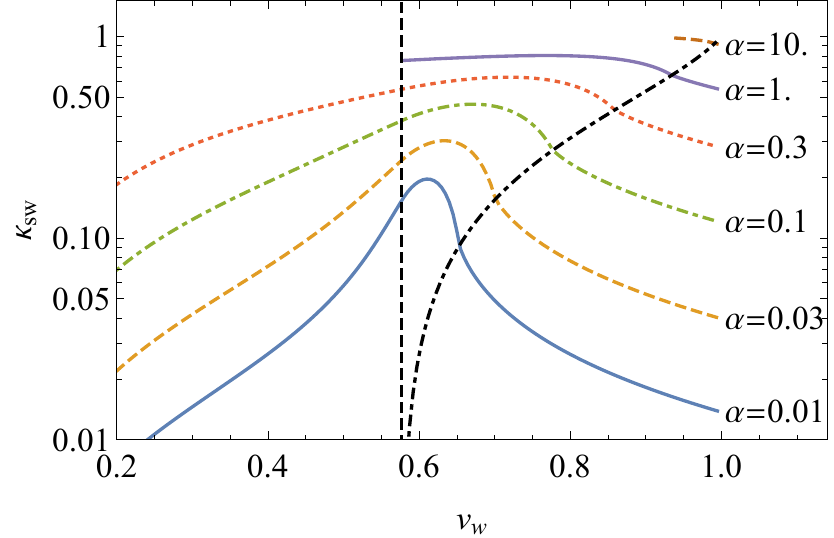}
\includegraphics[width=0.49\textwidth]{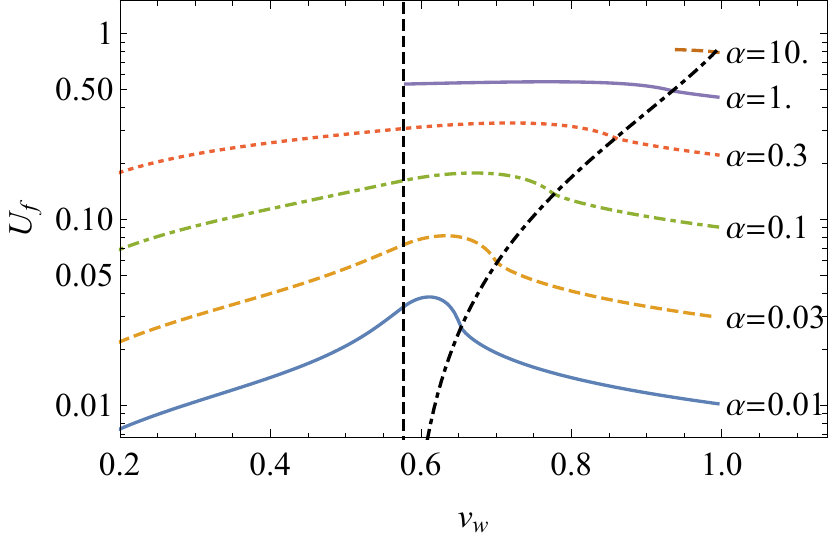}
\end{center}
\vspace{-5mm}
\caption{\it Efficiency of energy transfer into bulk fluid motion $\kappa_{\rm sw}$ (left panel) and RMS fluid velocity $\bar{U}_f$ (right panel) for a range of transition strengths $\alpha$ as functions of the wall velocity $\xi_{\rm w}$. The black dashed vertical line shows the boundary between deflagration (for $v_w<c_s$) and hybrid (for $v_w>c_s$) fluid profile solutions while the dot-dashed black line show the boundary between deflagration (for $\alpha<\alpha_{\rm det}^{\rm max}$) and detonation (for $\alpha<\alpha_{\rm det}^{\rm max}$) solutions (see Eq.~(\ref{eq:alpha_max_deto}) in Appendix~\ref{sec:Hydro_Appendix}).}
\label{fig:valocities_and_kappas}
\end{figure}

% \end{figure}
%
\noindent
A related quantity, which will also prove crucial for GW emission, is the root-mean-square (RMS) four-velocity of the plasma 
$\bar{U}_f$~\cite{Hindmarsh:2015qta,Hindmarsh:2017gnf,Hindmarsh:2019phv} (see also~\cite{Caprini:2019egz}), defined by
\be \label{eq:plasma_rms_velocity}
\bar{U}^2_f = \frac{1}{\overline{w}\, V} \int_V  T^{ii}\, d^3x =\frac{3}{\overline{w}\, v_w^3} \int w \,\xi^2 \frac{v^2}{1-v^2} d\xi \, ,
\ee
{where $V$ is the averaging volume and $\overline{w}$ is the average total enthalpy density in the symmetric phase. In the last step of~\eqref{eq:plasma_rms_velocity} we have particularised to a single expanding bubble, but the definition of $\bar{U}^2_f$ is more general, applicable to a system of many overlapping fluid shells at the end of the phase transition.\footnote{The results of numerical lattice simulations show that $\bar{U}^2_f$ from a single expanding bubble very closely matches the results from an ensemble of overlapping bubbles during gravitational wave production at the end of the phase transition~\cite{Hindmarsh:2015qta,Hindmarsh:2017gnf}.}}  
Based on the fact that $\overline{w} = \overline{e} + \overline{p}$ (with $\overline{p}$ and $\overline{e}$, respectively, the average pressure and average total energy in the symmetric phase), and thus
\be
\overline{w} =  (1 + \overline{p}/\overline{e})\, \overline{e} =  (1 + \overline{p}/\overline{e})\, (1+\alpha)\,\rho_{\mathrm{R}} \simeq 4/3\,(1+\alpha) \,\rho_{\mathrm{R}} \, ,
\ee
where in the last step we have used the fact that $p/e = 1/3$ for a relativistic perfect fluid,
one obtains a very simple approximation to the RMS fluid velocity~\cite{Hindmarsh:2017gnf}
\begin{equation}
    \bar{U}_f \simeq \sqrt{\frac{3}{4} \frac{\kappa \alpha}{1 + \alpha}} \, .
\end{equation}
We show the efficiency $\kappa$ and RMS fluid velocity for several values of $\alpha$ in Fig.~\ref{fig:valocities_and_kappas}, as functions of the wall velocity $v_w$ .

% \begin{figure}[h]    
% \begin{center}
% \includegraphics[width=0.65\textwidth]{kappaplot.pdf}\\
% \hspace{-0.5cm}\includegraphics[width=0.68\textwidth]{Ufapproxplot.pdf}\\ 
% \includegraphics[width=0.65\textwidth]{Ufplot.pdf}
% \end{center}
% \vspace{-5mm}
% \caption{\it Efficiency of energy transfer into bulk fluid motion $\kappa_{\rm sw}$ as well as the RMS fluid velocity $U_f$ and its simple approximation $\sqrt{\frac{3}{4}\frac{\kappa_{\rm sw} \alpha}{1+\alpha}}$ for a range of transition strengths $\alpha$ as a function of the wall velocity $\xi_{\rm w}$}.
% \label{fig:valocities_and_kappas2}
% \end{figure} 

At the end of the phase transition, the collisions of the bubbles result in the merging of fluid shells accompanying the respective bubbles, giving rise to a stage of linear fluid evolution (of overlapping fluid shells) after the transition, %known as the acoustic phase, 
during which GWs are sourced by the sound waves that develop in the plasma~\cite{Hindmarsh:2013xza,Hindmarsh:2015qta,Hindmarsh:2016lnk,Hindmarsh:2017gnf,Hindmarsh:2019phv} (see Section~\ref{sec:GWsignals}).
The timescale for shock formation $\tau_{\rm sh}$ in the plasma~\cite{Pen:2015qta} gives the duration of the linear fluid evolution period, after which the GW production from sound waves 
shuts off~\cite{Hindmarsh:2017gnf}. This makes $\tau_{\rm sh}$ a key quantity for GW generation, as it determines how long the plasma sound waves are active as a GW source. It can 
be estimated as~\cite{Pen:2015qta,Hindmarsh:2017gnf}
\be
\label{eq:tau_shock}
\tau_{\rm sh} \sim L_f/\bar{U}_f\, ,
\ee
where $L_f$ is the characteristic length of the fluid flow. The comparison of $\tau_{\rm sh}$ with the 
Hubble scale allows us to check whether the linear evolution of the fluid flow and the GW sources associated with it can be considered long-lasting ($H\, \tau_{\rm sh} \gtrsim 1$), which has a crucial impact on the GW generation from the phase transition and constitutes the main focus of this work.

The characteristic fluid length $L_f$ is in general given approximately by the mean bubble separation at percolation $R_*$, such that 
\be \label{eq:RHoverUf}
H \,\tau_{\rm sh} \sim \frac{H\,R_*}{\bar{U}_f}\approx \frac{(8\pi)^\frac13 \, {\rm Max}(v_w,c_s)}{\bar{U}_f}\left(\frac{\beta}{H}\right)^{-1} \, ,
\ee
%
%This is simply given by the ratio of the typical size of the bubbles at percolation to the RMS fluid velocity $ R_*/\bar{U}_f$~\cite{Hindmarsh:2017gnf,Ellis:2018mja}. 
%Assuming that we know the velocity of the bubble wall $v_w$ we can conveniently convert it into
where in the last step we have used the relation between $R_*$ and the duration of the phase 
transition~\cite{Caprini:2019pxz}. 
Recalling that we need $\alpha$ and $v_w$ to compute $\bar{U}_f$, it becomes apparent that 
for a certain bubble wall velocity, we can easily express $H \,\tau_{\rm sh}$ as a function of 
$\alpha$ and $\frac{\beta}{H}$, and so find the boundary between long-lasting 
($H \,\tau_{\rm sh} \gtrsim 1$) and short-lasting ($H \,\tau_{\rm sh} < 1$) plasma sound waves in the 
$(\alpha, \beta/H)$ parameter space.

%In all the subsequent figures we show the crucial boundary $HR_*/\bar{U}_f=1$ above which the flow quickly becomes non-linear and sound waves are no longer a long-lasting source of GWs. 

%%%%%%%%%%%%%%%%%%%%%%%%%%%%%%%%%%%%%%%%%%%%%%%%%%%%%%%%%%%%%%%%%%%%%%%%%%%%%
%%%%%%%%%%%%%%%%%%%%%%%%%%%%%%%%%%%%%%%%%%%%%%%%%%%%%%%%%%%%%%%%%%%%%%%%%%%%%
%%%%%%%%%%%%%%%%%%%%%%%%%%%%%%%%%%%%%%%%%%%%%%%%%%%%%%%%%%%%%%%%%%%%%%%%%%%%%

%The turbulence timescale, both for appearance and decay, is the shock appearance or eddy turn-over time

%%%%%%%%%%%%%%%%%%%%%%%%%%%%%%%%%%%%%%%%%%%%%%%%%%%%%%%%%%%%%%%%%%%%%%%%%%%%%%%%%%%%%%%%%
%%%%%%%%%%%%%%%%%%%%%%%%%%%%%%%%%%%%%%%%%%%%%%%%%%%%%%%%%%%%%%%%%%%%%%%%%%%%%%%%%%%%%%%%%
%%%%%%%%%%%%%%%%%%%%%%%%%%%%%%%%%%%%%%%%%%%%%%%%%%%%%%%%%%%%%%%%%%%%%%%%%%%%%%%%%%%%%%%%%

\section{Correlation between strength and duration of the phase transition: \\ Thin-wall scenario} 
\label{sec:thinwall}

As already pointed out at the end of Section~\ref{sec:definitions}, the strength and the duration of a cosmological first-order phase transition are not fully independent parameters, but rather are correlated for particle physics models. 
Our goal in this work is to discuss generic properties of the phase transition, rather than to study a specific example. From an analytical perspective, we need to make certain approximations in order to do so. 

\vspace{1mm}

In this Section 
we discuss the so-called thin-wall limit~\cite{Coleman:1977py}, in which the  bubble profile is approximated by just two field values, one inside and the other outside the bubble:
\be 
\phi_{\rm tw}=
\begin{cases}
\phi_{\rm t} \quad {\rm for} \quad r < R_0 \, , \\
\phi_{\rm f} \quad {\rm for} \quad r > R_0 \, ,
\end{cases}
\ee
where $R_0$ is the radius of the bubble. % and the field values $\phi_{\rm t,f}$ correspond to the field values in the true and false vacuum, respectively.
The key part of this solution is finding the radius of the bubble at nucleation, which turns out to be (see, e.g.,~\cite{Darme:2017wvu})
\be
R_0=\frac{3\sigma}{\Delta V}\, ,
\ee
just as in the four-dimensional case where, as usual,
\be
\sigma=\int_{\phi_{\rm f}}^{\phi_{\rm t}} d\phi \sqrt{2 V(\phi)-V_{\rm t}} \, ,
\ee
and we recall that we set the constant term in the potential such that $V(\phi_{\rm f})=0$ (hence the extra $-V_{\rm t}$ term in the square root). Using the thin-wall approximation
we can simply write the action~\eqref{eq:bounce_action_expanded} as 
\be \label{eq:S3tw}
S_3=-2 \times 4\pi \int dr r^2  V(\phi)=\frac{4\pi}{3}\times 2 R_0^3 \,\Delta V=\frac{8\pi}{3} \,\frac{27 \sigma^3}{(\Delta V)^2} = 72\,  \frac{\sigma^3 }{\alpha^2 \rho_{\rm r}^2}=\frac{72\, \sigma^3}{\alpha^2 T^8 \xi_g^4}\, ,
\ee
where $\xi_g=\sqrt{30/\pi^2 g_*}$ and $g_*$ is the number of relativistic degrees of freedom at the time of 
the phase transition (in the phase $\phi_{\rm f}$). In potentials where the thin-wall solution is a good approximation, $\Delta V$ is only a small correction term in the potential. This means that $\sigma$ is modified by only a negligible additive constant when $\alpha$ changes.
Thus, {in order to compute $\beta$ from~\eqref{eq:S3tw}} we need to include only the explicit polynomial dependence on the temperature, which implies
\be
\frac{\beta}{H}=T\frac{d}{dT}\frac{S_3}{T}\propto \frac{S_3}{T} \propto \alpha^{-2}\, ,
\ee
yielding a simple polynomial dependence of $\beta/H$ on $\alpha$. 
{Remembering also that} the Hubble expansion rate is
\be
H^2=\frac{1}{3 M_p^2}\left( \rho_{\rm R} +\rho_V \right)=\frac{1}{3 M_p^2}\left( \frac{T^4}{\xi_g^2} +\Delta V \right) \, ,
\ee
we can rewrite the nucleation condition~\eqref{eq:nucleation_temperature} as
\be
\label{eq:nucleation_crit}
1\approx\frac{\Gamma}{H^4} \Longrightarrow \frac{S_3}{T} \approx -\log \left( \frac{1}{9M_p^4}\frac{T^4}{\xi_g^4} \right) \approx 167-4\log (T/{\rm GeV}) \, ,
\ee
where we have fixed for simplicity $g_* = 100$ (approximating the number of relativistic d.o.f. in the early Universe around the electroweak scale).
Eq.~(\ref{eq:nucleation_crit}) allows us to fix the value of $S_3/T$ and consequently $\beta/H$ at the nucleation temperature. However we must still find the value of $\alpha$ at that reference temperature, so this simplified approach only yields the value of $\beta/H$ as a function of $\alpha$ up to a multiplicative constant.

\begin{figure}[t]
%\captionsetup{width=.8\linewidth}
\begin{center}
\includegraphics[width=0.75\textwidth]{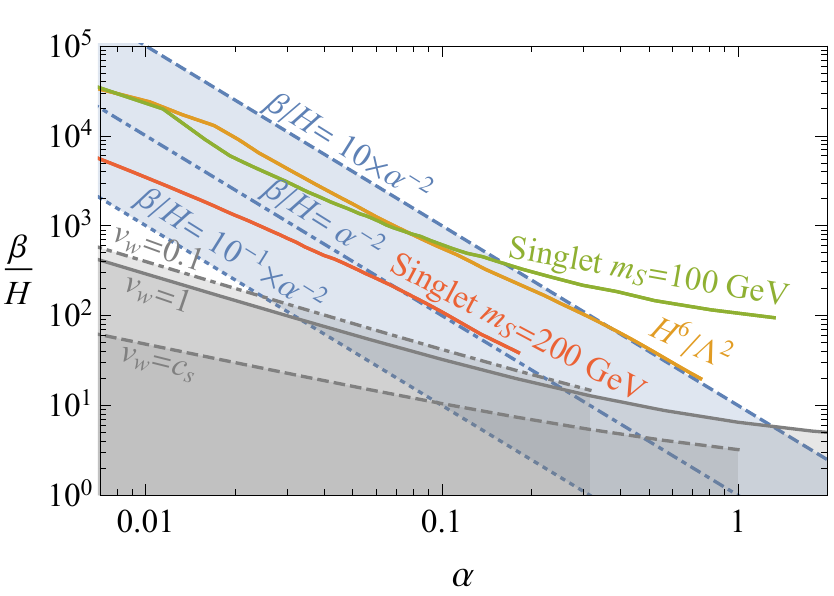}
\end{center}
\vspace{-5mm}
\caption{\it $\beta/H$ as a function of $\alpha$ for the 
simple thin-wall approximation, with the multiplicative constant (see text for details) set to $10$ (dotted blue), $1$ (dot-dashed blue) and $0.1$ (dashed blue). Also shown are the numerical results for a singlet scalar extension of the SM with 
the singlet mass $m_s = 100$ GeV (solid green) and $m_s = 200$ GeV (solid red), as well as for the SM extended by a non-renormalizable term $\propto H^6$ (solid yellow). The grey regions correspond to $H \tau_{\rm sh} > 1$ for different values of the bubble wall velocity.
}
\label{fig:betaHofalphaplot}
\end{figure}

\vspace{2mm}

The results for the dependence of $\beta/H$ on $\alpha$ in the thin-wall limit for several values of the unknown multiplicative constant are shown in Fig.~\ref{fig:betaHofalphaplot}.
For comparison we also show 
realistic results from two BSM scenarios, namely the SM extended with an $H^6$ non-renormalizable operator and the singlet scalar extension of the SM (details can be found in~\cite{Ellis:2018mja,Ellis:2019oqb}).
As shown in Fig.~\ref{fig:betaHofalphaplot}, the simple thin-wall approximation captures reasonably well the trend found in realistic scenarios. 
Fig.~\ref{fig:betaHofalphaplot} also shows the approximate location of the boundary $H \tau_{\rm sh} = 1$ between short- and long-lasting plasma sound waves, for different values of the bubble wall velocity $v_w$ up to the maximum value of $\alpha$ allowed for each value of $v_w$. The results from Fig.~\ref{fig:betaHofalphaplot} already indicate that for weaker phase transitions (smaller $\alpha$) the lifetime of the sound wave GW source compared to a Hubble time is in general shorter. At the same time, the long-lifetime region $H \tau_{\rm sh} \gtrsim 1$ is approached only for very strong transitions (particularly if $(\beta/H)\,\alpha^2 > 1$ in the thin-wall approximation). In the next Section we analyze this in more detail away from the thin-wall limit. 

%%%%%%%%%%%%%%%%%%%%%%%%%%%%%%%%%%%%%%%%%%%%%%%%%%%%%%%%%%%%%%%%%%%%%%%%%%%
%%%%%%%%%%%%%%%%%%%%%%%%%%%%%%%%%%%%%%%%%%%%%%%%%%%%%%%%%%%%%%%%%%%%%%%%%%%
%%%%%%%%%%%%%%%%%%%%%%%%%%%%%%%%%%%%%%%%%%%%%%%%%%%%%%%%%%%%%%%%%%%%%%%%%%%
\section{General results for polynomial \& Coleman-Weinberg-like potentials} \label{sec:genral_polynomial}

We now analyze the possible lifetime of the sound-wave source for some generic scalar potentials. We first study the case of a polynomial potential without a potential barrier between the false and true vacua at $T = 0$, corresponding to a purely thermal phase transition. Then we investigate the case of a polynomial potential with a non-zero barrier at $T = 0$, and finally the case of a Coleman-Weinberg-like (classically scale-invariant) potential. In the latter two cases, significant supercooling during the phase transitions is possible, which could lead to a breakdown of the relation
$\Gamma \propto e^{-\beta t}$ from Eq.~(\eqref{eq:beta_Taylor}). This is typically the case for slow phase transitions, with $\beta/H \lesssim 10$.
In order to deal with this problem in our analysis, throughout this work we use the refined analysis of percolation and phase transition completion described in detail in~\cite{Ellis:2018mja} and discussed here in Appendix~\ref{sec:percolation}.

%%%%%%%%%%%%%%%%%%%%%%%%%%%%%%%%%%%%%%%%%%%%%%%%%%%%%%%%%%%%%%%%%%%%%%%%%%%%%%%%
\subsection{Purely thermal transition: no $T=0$ potential barrier} 
\label{sec:thermal_polynomial} 

We consider first the simplest renormalisable scalar potential, which takes the form 
\be \label{eq:renormalisableV_total}
\begin{split}
V & =m^2 \phi^2- a\, \phi^3+  \lambda \phi^4 \, ,
\end{split}
\ee
where $\lambda$, $a$ and $m^{2}$ are parameters that may in general depend on the temperature $T$. For this simple potential, there is an accurate semi-analytical approximation to the tunnelling action~\cite{Adams:1993zs}:
 \be \label{eq:approx_generic_poly_action}
\frac{S_3}{T}=\frac{a}{T \lambda ^{3/2}} \frac{8 \pi \sqrt{\delta } \left({\beta_1} \delta +\beta_2 \delta ^2+\beta_3 \delta ^3\right)}{81 (2-\delta )^2 },
\; \; {\rm where} \; \;
\delta \equiv \frac{8\lambda m^2}{a^2} \, ,
\ee
with $\beta_1 = 8.2938$, $\beta_2 = -5.5330$ and $\beta_3 = 0.8180$. 

\vspace{1mm}

Using the high-temperature expansion for the 1-loop finite-temperature effective potential (see~\cite{Quiros:1999jp} for details), we can use the simple potential from~\eqref{eq:renormalisableV_total} to approximate realistic particle physics scalar potentials:
\be \label{eq:thermalbarrierV}
V(\phi,T)  = \frac{g_{m^2}}{24}\left(T^2-T_0^2\right) \phi^2  -\frac{g_m}{12\pi} T \phi^3 +  \lambda \phi^4\, ,
\ee
where we consider $T_0^2 > 0$, in which case $T_0$ corresponds to the temperature below which the potential barrier disappears and the false vacuum $\phi_{\rm f} = 0$ becomes unstable. This results in a purely thermal phase transition.

For realistic particle physics models, the dimensionless constant $g_{m^2}$ in~\eqref{eq:thermalbarrierV} is related to the squared-mass shifts at the bubble wall for all the particles coupled to $\phi$, due to the change in vacuum expectation value $\Delta\phi$ across the phase boundary. This corresponds to the number of degrees of freedom coupled to $\phi$ weighted by their couplings squared ($g_f$ for fermions and $g_b$ for bosons). The interpretation of 
$g_{m}$ is similar except that now only linear mass shifts in bosonic degrees of freedom contribute, weighted by their couplings $g_b$ squared:
\be
\begin{split} \label{eq:thermalpotparams}
g_{m^2} & \equiv \; \frac{1}{(\Delta \phi)^2} \left(
\sum_b N_b \Delta m_b^2+\frac{1}{2}\sum_f N_f \Delta m_f^2 \right) =\sum_b  N_b \,g_b^2+\frac{1}{2}\sum_f  N_f\, g_f^2.
\\
g_m & \equiv \; \frac{1}{\Delta \phi} \left( \sum_b N_b\,g_b^2 \Delta m_b \right)=\sum_b N_b\, g^3_b \, .
\end{split}
\ee
Combining~\eqref{eq:thermalbarrierV} with~\eqref{eq:approx_generic_poly_action} and~\eqref{eq:renormalisableV_total}, after some algebra we can express the $\alpha$ and $\beta/H$ parameters as simple analytical functions. The explicit results are a bit lengthy and we show them in Appendix~\ref{app:PTparameters}. 
\begin{figure}[t]
\begin{center}
\includegraphics[width=0.49\textwidth]{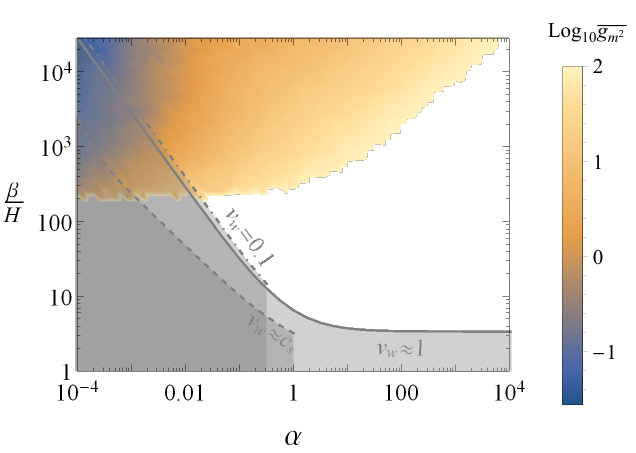}
\includegraphics[width=0.49\textwidth]{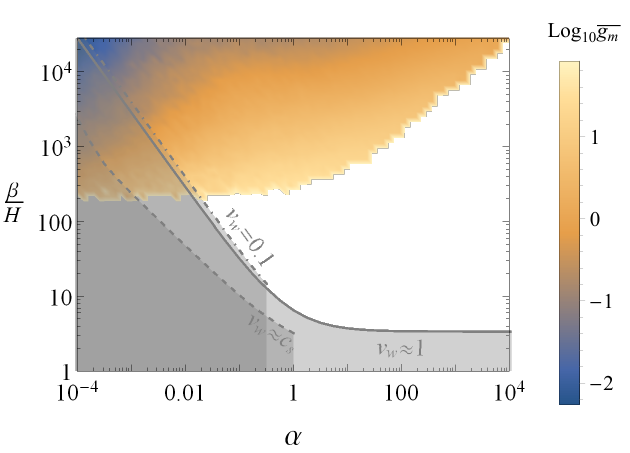}
\includegraphics[width=0.49\textwidth]{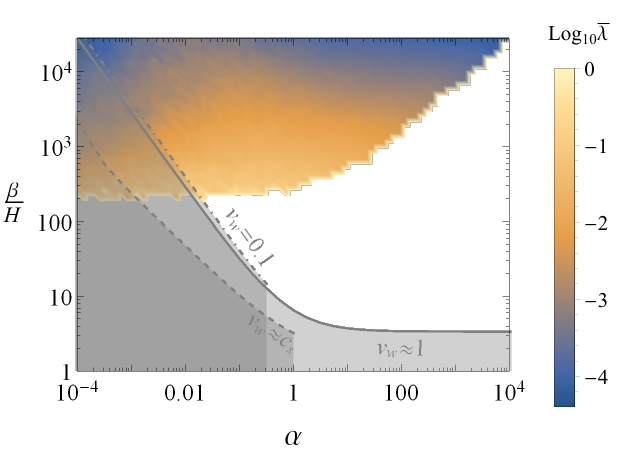}
\includegraphics[width=0.49\textwidth]{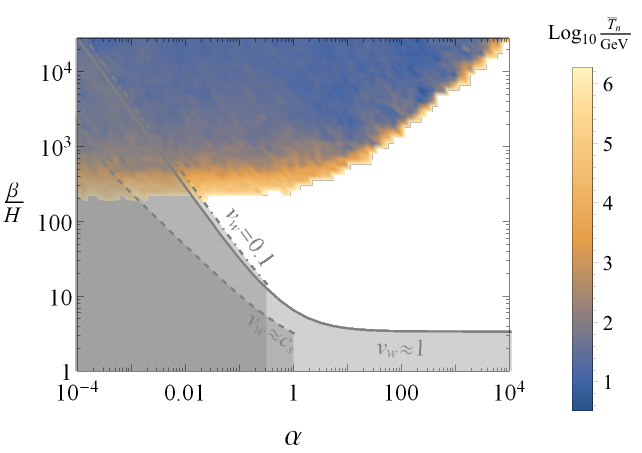}
\end{center}
\vspace{-5mm}
\caption{\it Regions of the $(\alpha, \beta/H)$ plane realised for the polynomial potential with a purely thermal barrier, Eq.~\eqref{eq:thermalbarrierV}. The colour shading indicates the geometric average of the scan for the values of $g_{m^2}$ (upper left), 
$g_m$ (upper right) %(dashed black lines indicate the SM values $g_m^2=4$ and  $g_m=0.36$ ) 
and $\lambda$ (lower left) in the potential~\eqref{eq:thermalbarrierV}, as well as the nucleation temperature $T_n$ (lower right) for each point in the plane. The grey shading indicates the regions with long-lived sound-wave emission for different bubble wall velocities.
}
\label{fig:TBscan}
\end{figure} 
Our analysis then consists of finding the percolation temperature as described in Appendix~\ref{sec:percolation} and using the explicit results from Appendix~\ref{app:PTparameters} to obtain the key transition parameters at percolation. This remarkable simplification, compared to the standard search for a bounce solution, allows us to perform a simple scan over the parameter space to show the landscape of possible results, shown explicitly in Fig.~\ref{fig:TBscan}. The ranges chosen for the parameter scan (with a logarithmic prior) are
\be
\begin{split}
10^{-4} \leq &\, g_{m^2} \leq 10^2  \, ,  \\
10^{-4} \leq &\, g_m \leq g_{m^2} \, , \\
10^{-6} \leq &\,  \lambda \leq 10 \, , \\
10^{-3}\, {\rm GeV} \leq &\, T_0 \leq 10^5 \, {\rm GeV} \, .
\end{split}
\ee
Our chosen scan upper limit for $g_{m^2}$ is motivated from~\eqref{eq:thermalpotparams}, by considering $N_b + N_f/2 \lesssim 100$ and $g_b, \,g_f \lesssim 1$. We stress that close to these boundaries perturbation theory ceases to be reliable, and our results should there be interpreted as indicative at best. The condition $g_{m} < g_{m^2}$ in our scan is also motivated by requiring $g_b, \,g_f \lesssim 1$ (recall Eq.~\eqref{eq:thermalpotparams}). 
%Finally, from eq.~\eqref{eq:alpha_TB} in Appendix~\ref{app:PTparameters} a lower bound exists on the value of $\lambda$ to guarantee $\alpha > 0$, given by $\lambda >  $ (which is also implemented in our scan). 

We have verified the accuracy of our scanning method by comparing the results obtained for a subset of points in the scan against a full calculation based on finding the bounce solution (as explained e.g. in~\cite{Ellis:2018mja,Ellis:2019oqb}), finding very good agreement for all the verified points (the difference comes mostly from approximating the transition temperature by the nucleation temperature instead of the percolation temperature). We have also verified the fit to the bounce action given in~\cite{Adams:1993zs}.\footnote{Specifically, while in any particular transition the values of $\beta_i$ parameters in~\eqref{eq:approx_generic_poly_action} can change slightly, this has a negligible impact on the resulting nucleation or percolation temperature, as well as on $\alpha$ and $\beta/H$.}
%, finding good agreement.

%
\begin{figure}[h]
\begin{center}
\includegraphics[width=0.495\textwidth]{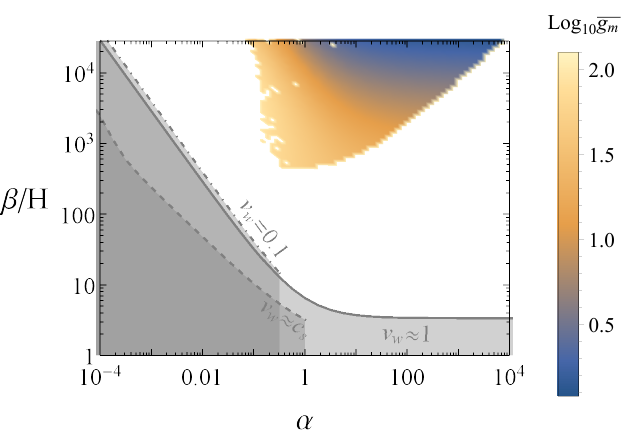}
\includegraphics[width=0.495\textwidth]{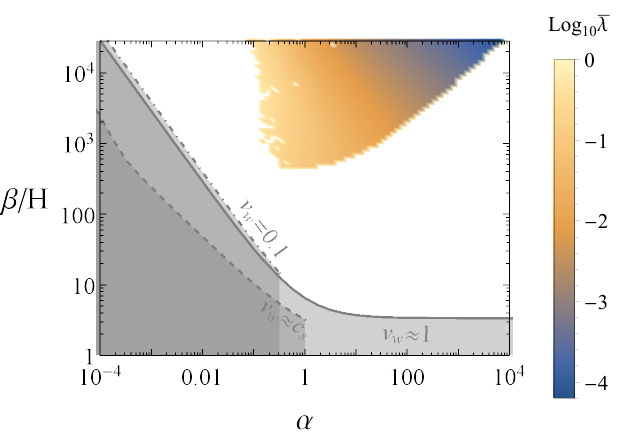}
\end{center}
\vspace{-5mm}
\caption{\it Regions of the $(\alpha, \beta/H)$ plane realised for the polynomial potential with a purely thermal barrier, Eq.~\eqref{eq:thermalbarrierV}, for fixed $g_{m^2} = 100$ and $T_0 = 100$ GeV. The colour shading indicates the geometric average of the scan for the values of $g_{m}$ (left) and $\lambda$ (right) for each point in the plane. The grey shading indicates the regions with long-lived sound-wave emission for different bubble wall velocities.
}
\label{fig:TBscan_Fixed}
\end{figure} 

From the results in Fig.~\ref{fig:TBscan} we see immediately that there exists an approximate lower bound on the value of $\beta/H$ (whose specific value depends on the upper and lower limits of the scan parameters $g_{m^2}, g_{m}$ and $\lambda$) when the transition is purely thermal and the potential barrier between vacua disappears at some finite temperature $T_0$. At the same time, 
Fig.~\ref{fig:TBscan} shows that transition strength values $\alpha \gg 1$ are compatible with successful percolation even in this case, provided that the nucleation rate is fast enough, which is possible in regions of the parameter space where $\beta/H$ is sufficiently large. This happens for large values of $g_{m^2}$ together with very small values of both $g_{m}$ and $\lambda$ in~\eqref{eq:thermalbarrierV}, as shown in Fig.~\ref{fig:TBscan_Fixed}. 
However, we note that in this corner of the parameter space the potential~\eqref{eq:thermalbarrierV} is not expected to capture well the behaviour of realistic particle physics models: Eq.~\eqref{eq:thermalbarrierV} neglects both thermal and quantum corrections to $\lambda$, and it is precisely for very small tree-level values of $\lambda$ that they become important, increasing the overall size of the quartic term in Eq.~\eqref{eq:thermalbarrierV} (and adding a temperature dependence to it). As a result, large values of $\alpha$ in scenarios with a purely thermal potential barrier between vacua are very difficult to obtain in general.

%%%%%%%%%%%%%%%%%%%%%%%%%%%%%%%%%%%%%%%%%%%%%%%%%%%%%%%%%%%%%%%%%%%%%%%%%%%%%%%%%%%%%%%%%%%
\subsection{Tree-level potential barrier at $T = 0$} 
\label{sec:treelevel_polynomial}
We turn in this Section to the discussion of polynomial potentials with a barrier between vacua that persists down to $T=0$,
as can occur in multi-scalar models where the transition proceeds with non-zero vev's of new scalars, and in models
with a dark sector~\cite{Curtin:2014jma,Beniwal:2017eik,Baldes:2017rcu,Breitbach:2018ddu}. In comparison to the analysis of the previous Section, we now
introduce a potential barrier at tree-level via a term $A \phi^3$, and for simplicity assume that it dominates over the thermal cubic term in~\eqref{eq:thermalbarrierV},~$A \gg g_{m} T_0$, so that we can 
neglect the latter. The resulting scalar potential reads
\be \label{eq:zeroTbarrierV}
V(\phi,T) = \frac{g_{m^2}}{24}\left(T^2-T_0^2\right) \phi^2 - A \phi^3 + \lambda \phi^4\, ,
\ee
where we again relate $g_{m^2}$ to the mass shifts across the bubble wall, see~\eqref{eq:thermalpotparams}.  Also, now $T_0^2 < 0$ is allowed so that the false vacuum $\phi_{\rm f} = 0$ can persist as a (local) minimum down to $T=0$. 
Combining~\eqref{eq:approx_generic_poly_action} and~\eqref{eq:zeroTbarrierV} we can express the parameters $\alpha$ and $\beta/H$ in analytical form, as shown explicitly in Appendix~\ref{app:zeroTbarrier}. 
\begin{figure}[t]
\begin{center}
\includegraphics[width=0.49\textwidth]{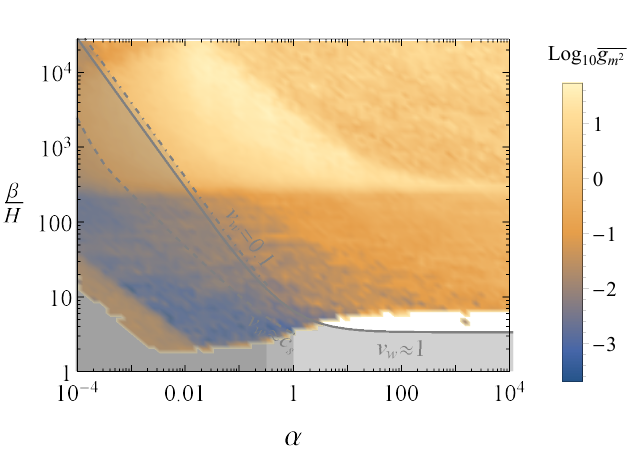}
\includegraphics[width=0.49\textwidth]{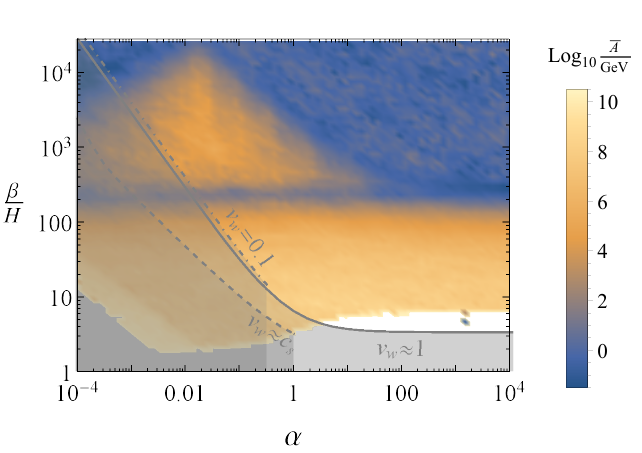}
\includegraphics[width=0.49\textwidth]{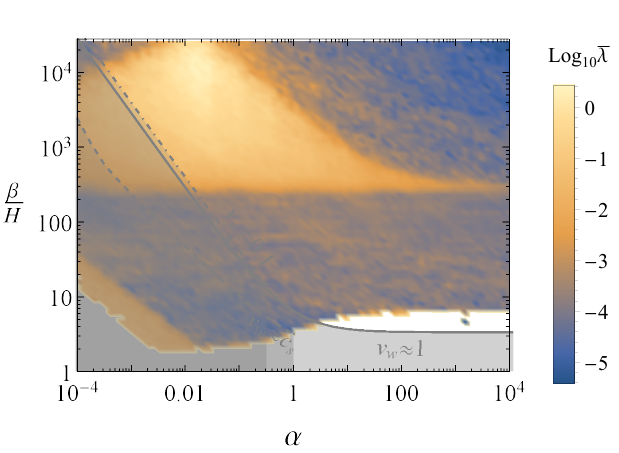}
\includegraphics[width=0.49\textwidth]{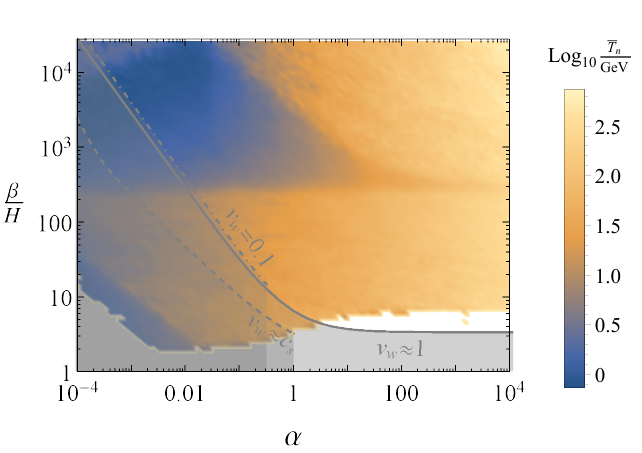}
\end{center}
\vspace{-5mm}
\caption{\it Allowed regions of the $(\alpha, \beta/H)$ plane for the potential~\eqref{eq:zeroTbarrierV}. The colour shading indicates the geometric average of the  values of $g_m^2$ (upper left), $A/{\rm GeV}$ (upper right) and $\lambda$ (lower left) in the potential~\eqref{eq:zeroTbarrierV}, as well as the nucleation temperature $T_n/{\rm GeV}$ (lower right) for each point in the plane. The grey shading indicates the regions with long-lived sound-waves for different bubble wall velocities.
}
\label{fig:ZTBscan}
\end{figure} 
We again perform a scan over the parameter space, within the ranges
\be
\begin{split}\label{eq:zeroTscan}
10^{-6} \leq & g_{m^2} \leq 10^2 \, ,   \\
10^{-3}\, {\rm GeV}  \leq & A \leq 10^{10} \, {\rm GeV} \, , \\
10^{-8} \leq &  \lambda \leq 10 \, , \\
10^{-3}\, {\rm GeV} \leq & T_0 \leq 10^{5} \, {\rm GeV}
\end{split}
\ee
also considering for each point both signs for $T_0^2$ in ~\eqref{eq:zeroTbarrierV}. We show the results in Fig.~\ref{fig:ZTBscan}. 

Compared to the results of Section~\ref{sec:thermal_polynomial}, 
in this case there is a broader range of the parameter space that may yield long-lived sound waves as a source of GWs, particularly since now low values of $\beta/H$ of $\mathcal{O}(10-100)$ become possible. Despite the fact that some of the results predict very low $\beta/H$, we emphasise again that for each of the points percolation has been verified following the prescription in Appendix~\ref{sec:percolation}. At the same time, the average bubble size $R_*$ in each of those cases proves to be significantly below the horizon size, even though a naive exponential nucleation prescription would suggest the contrary. Similarly to the case of a purely thermal potential barrier studied in the previous Section, at very low values of $\lambda$ below $10^{-4}-10^{-5}$, the potential~\eqref{eq:zeroTbarrierV} ceases to yield a good approximation to realistic scenarios, for which quantum and thermal corrections to $\lambda$ would be important in such regime.\footnote{As a result, the blue regions in Fig.~\ref{fig:ZTBscan} (lower left panel) should be interpreted as only indicative.}
From the upper left panel of Fig.~\ref{fig:ZTBscan}, we also infer that sound waves with a lifetime longer than a Hubble time tend to have small values of $g_{m^2}$, so as to create a large hierarchy between the potential squared-mass parameter $\mu^2 \equiv (g_{m^2}/24) \times T_0^2$ and the squared-temperature $T_0^2$.

%%%%%%%%%%%%%%%%%%%%%%%%%%%%%%%%%%%%%%%%%%%%%%%%%%%%%%%%%%%%%%%%%%%%%%%%%%%%%%%%%%%%%%%%%%%
%%%%%%%%%%%%%%%%%%%%%%%%%%%%%%%%%%%%%%%%%%%%%%%%%%%%%%%%%%%%%%%%%%%%%%%%%%%%%%%%%%%%%%%%%%%
%%%%%%%%%%%%%%%%%%%%%%%%%%%%%%%%%%%%%%%%%%%%%%%%%%%%%%%%%%%%%%%%%%%%%%%%%%%%%%%%%%%%%%%%%%% 
 
\subsection{Classically scale-invariant Coleman-Weinberg-like potential} \label{sec:scaleinvariant}

\begin{figure}[h]
\begin{center}
\includegraphics[width=0.49\textwidth]{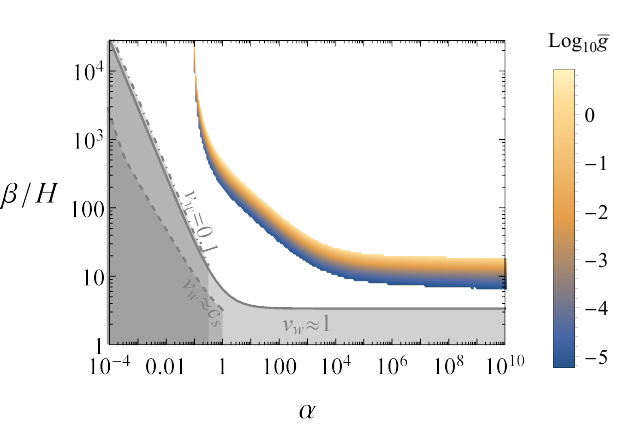}
\includegraphics[width=0.49\textwidth]{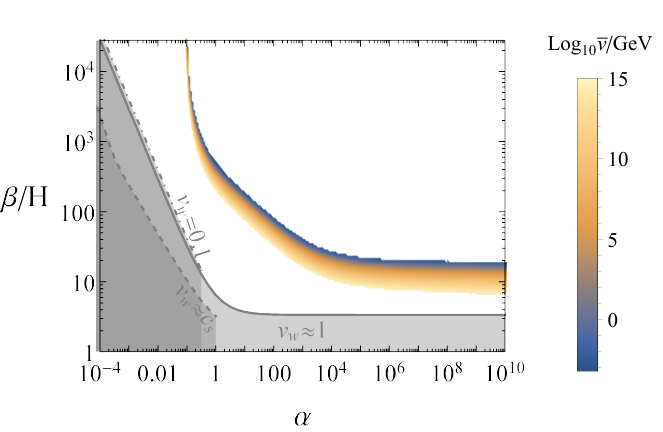}
\includegraphics[width=0.49\textwidth]{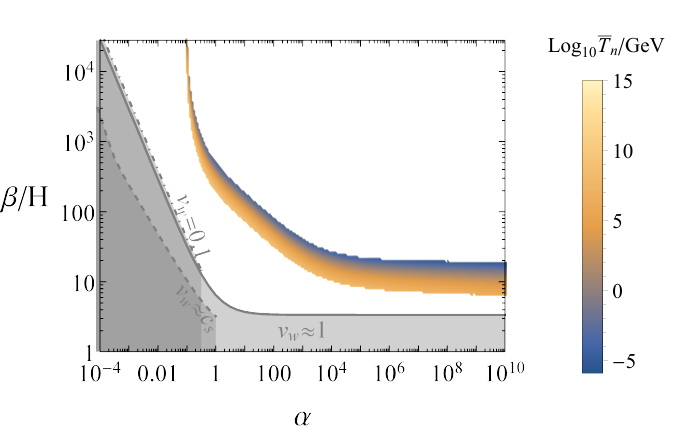}
\end{center}
\vspace{-5mm}
\caption{\it Allowed regions of the $(\alpha, \beta/H)$ plane for scale-invariant Coleman-Weinberg-like scalar potentials like~\eqref{eq:SIpotential}.
The colour shading indicates the geometric average of the  values of $g$ (upper left) and $v/{\rm GeV}$ (upper right), as well as the nucleation temperature $T_n/{\rm GeV}$ (lower right) for each point in the plane. The grey shading indicates the regions with long-lived sound-waves for different bubble wall velocities.
}
\label{fig:SIscan}
\end{figure}

We turn finally to a very different class of potentials arising in many extensions of the SM, namely (almost) scale-invariant Coleman-Weinberg-like potentials. A generic such potential can be written as~\cite{Adams:1993zs}
\be \label{eq:SIpotential0}
V(\phi, T)=\left( 2 B_1-B_2 \right)\sigma^2 T^2-B_1\phi^4+B_2\phi^4 \log\left(\frac{\phi}{\sigma}\right)\, .
\ee  
We again use an analytical approximation to the tunnelling action~\cite{Adams:1993zs}:
\be 
\frac{S_3}{T}=\frac{16\pi\sigma I^3}{3\,T\left(1-2\delta \right)^2} \sqrt{\frac{2}{B_2}}\left(2\delta \right)^{n_\mu}\left(1 +\mu_1 \delta + \mu_2 \delta^2 + \mu_3 \delta^3 \right)\, , \quad \delta=\frac{2B_1-B_2}{2B_2} \, ,
\ee
where $I=0.4199$, $n_\mu=0.557$, $\mu_1=4.27$, $\mu_2=-14.59$ and $\mu_3=12.09$. 
We set
\be
B_1=\frac{1}{2} \left(\frac{g^2 T^2}{\sigma^2}+\frac{3g^4}{4\pi^2} \right)\, , \quad B_2=\frac{3g^4}{4\pi^2}\, , \quad \sigma=v\, ,
\ee
so that
\be \label{eq:SIpotential}
V(\phi, T)= g^2 T^2 \phi^2 +\frac{3g^4}{4\pi^2} \phi^4 \left( \log \left(\frac{\phi^2}{v^2}\right)-\frac{1}{2}-\frac{g^2T^2}{2 v^2}   \right)\, , \quad \delta=\frac{2 \pi ^2 T^2}{3 g^2 v^2}\, ,
\ee
which is a reasonable approximation for potentials encountered in particle physics models featuring scale invariance (e.g.,~the $U(1)_{B-L}$ model studied in~\cite{Ellis:2019oqb}).  
We have again found analytical formulae for the phase transition parameters, the explicit results shown in Appendix~\ref{app:scaleinvariant}.
In this scenario there are only $2$ free parameters, namely the coupling $g$ and the $T=0$ scalar field vacuum expectation value $v$, which simply sets the scale of new physics for the model in question.
We perform a scan over them within the ranges
\be 
\begin{split}
10^{-8} <  g < 10\, ,  \quad
10^{-5} <  v/{\rm GeV} < 10^{15}\, ,
\end{split}    
\ee
in order to explore the available parameter space.
The results are shown in Fig.~\ref{fig:SIscan}. 
Within the allowed $(\alpha, \beta/H)$ region, we observe that decreasing values 
of $\beta/H$ are correlated with smaller values of $g$ and larger scales $v$. 
Values of $g$ similar to those in the SM lie well within the range of the scan, and give predictions for $\alpha$ and $\beta/H$ clearly outside the region where sound waves are long-lived. The same is true for all values of $v$ between ${\cal O}(\rm GeV)$ and a typical GUT scale. Altogether, Fig.~\ref{fig:SIscan}
highlights that for scale-invariant Coleman-Weinberg-like scalar potentials the sound wave GW source is only active for a small fraction of a Hubble time. 

%%%%%%%%%%%%%%%%%%%%%%%%%%%%%%%%%%%%%%%%%%%%%%%%%%%%%%%%%%%%%%%%%%%%%%%%%%%%%%%%%%%%%%%%%%%
%%%%%%%%%%%%%%%%%%%%%%%%%%%%%%%%%%%%%%%%%%%%%%%%%%%%%%%%%%%%%%%%%%%%%%%%%%%%%%%%%%%%%%%%%%%
%%%%%%%%%%%%%%%%%%%%%%%%%%%%%%%%%%%%%%%%%%%%%%%%%%%%%%%%%%%%%%%%%%%%%%%%%%%%%%%%%%%%%%%%%%%

%%%%%%%%%%%%%%%%%%%%%%%%%%%%%%%%%%%%%%%%%%%%%%%%%%%%%%%%%%%%%%%%%%%%%%%%%%%%%%%%%%%%%%%%%%%%%%%%%%%%%%%
%%%%%%%%%%%%%%%%%%%%%%%%%%%%%%%%%%%%%%%%%%%%%%%%%%%%%%%%%%%%%%%%%%%%%%%%%%%%%%%%%%%%%%%%%%%%%%%%%%%%%%%
%%%%%%%%%%%%%%%%%%%%%%%%%%%%%%%%%%%%%%%%%%%%%%%%%%%%%%%%%%%%%%%%%%%%%%%%%%%%%%%%%%%%%%%%%%%%%%%%%%%%%%%

\vspace{2mm}

\section{Gravitational wave signals}
\label{sec:GWsignals}

\subsection{Energy budget}

We now review the energy budget of the phase transition~\cite{Espinosa:2010hh} (see also~\cite{Ellis:2019oqb}), in order to discuss the implications of the results from the previous Sections for the production of GWs at the end of the phase transition. First, to compute the fraction of the energy used to accelerate the bubble walls we need the friction acting on the wall, which may be conveniently expressed through the parameters $\alpha_\infty$ and $\alpha_{\rm eq}$, see~\cite{Ellis:2019oqb}.
In the case of a purely thermal transition, both $\alpha_\infty$ and $\alpha_{\rm eq}$ depend directly on the parameters of our scan, and may be readily expressed as:
%
%we already have all the necessary information as the sums over degrees of freedom and their couplings enter also as parameters in the potential~\eqref{eq:thermalbarrierV}
\be 
\label{alpha_inf_eq_1}
\begin{split}
\alpha_\infty & =\frac{\Delta m^2 T^2}{24 \rho_R}=\frac{1}{24\frac{\pi^2}{30}g_*}\left(\frac{\Delta \phi}{T}\right)^2 
g_{m^2} \, ,%\left( \sum_b  N_b g_b^2+\frac{1}{2}\sum_f  N_f g_f^2 \right) \, ,
\\
\alpha_{\rm eq} & =\frac{g^2\Delta m_V T^3}{\rho_R} \sim \frac{1}{\frac{\pi^2}{30}g_*}\, \frac{\Delta \phi}{T}\, g_m \, . %\sum_b N_b g^3_b \, .
\end{split}
\ee
We note that we have conservatively assumed in~\eqref{alpha_inf_eq_1} that  
all bosons (as included in $g_m$) contribute to $\alpha_{\rm eq}$, whereas in practice only gauge bosons do~\cite{Bodeker:2017cim} so the value of $\alpha_{\rm eq}$ may also be somewhat smaller. 
The expressions in~\eqref{alpha_inf_eq_1} also apply to the tree-level $T=0$ potential barrier scenario discussed in Section~\ref{sec:treelevel_polynomial}, though in that case our scan does not include $g_m$. In the following we assume for definiteness 
a value $g_m \sim 0.1\, g_{m^2}$ (to emulate the suppression w.r.t. $g_{m^2}$ from the lack of fermion contributions together with an additional power of the appropriate coupling)
when discussing GW production in this scenario (bearing in mind that a departure of $g_m$ from this value would influence our GW results).
Finally, in the scale-invariant Coleman-Weinberg-like scenario discussed in 
Section~\ref{sec:scaleinvariant},
$\alpha_\infty$ can be directly obtained from the parameters $g$ and $v$, and 
we assume the minimal value of $\alpha_{\rm eq}$ given by a model of $U(1)_{B-L}$ gauge symmetry breaking (see e.g.~\cite{Ellis:2019oqb}).
%case we assume a minimal particle content of 3 massive gauge degrees of freedom, as would be the case for a $U(1)$ gauge group.
A summary of the expressions for $\alpha_\infty$ and $\alpha_{\rm eq}$ for the three generic types of phase transitions is given in Table~\ref{tab:frictioncoefficients}.
\begin{table}[h]
\centering
%%%%%%%%%%%%%%%%%%%%%%%%%%%%%%%%%%%%%%%%%%%%%%%%%%%%%%%%%%%%%%%%%%%%%%%%%%%%%
\begin{tabular}{ |c|c|c| } 
\hline
Model & $\alpha_\infty$ & $\alpha_{\rm eq}$  
\\ \hline & & \\[-2ex]
Thermal barrier & $\frac{5\, g_{m^2}}{4\,\pi^2 g_*}\left(\frac{\Delta \phi}{T}\right)^2$ & $\frac{30\, g_m}{\pi^2 g_*}\left(\frac{\Delta \phi}{T}\right)$  
\\[2ex] \hline & & \\[-2ex]
Zero-temperature barrier & $\frac{5\, g_{m^2}}{4\,\pi^2 g_*}\left(\frac{\Delta \phi}{T}\right)^2$ & $\frac{30\, g_m}{\pi^2 g_*}\left(\frac{\Delta \phi}{T}\right)$  
\\[2ex] \hline & & \\[-2ex]
Scale-invariant (CW-like) & $\frac{15\,g^2}{\pi^4 g_*}\left(\frac{v}{T}\right)^2$ & $\frac{180 \, g^3}{\pi^3 g_*}\left(\frac{v}{T}\right)$  \\[2ex]
\hline
\end{tabular}
\caption{\it Expressions for $\alpha_\infty$ and $\alpha_{\rm eq}$ in the scenarios from Sections~\ref{sec:thermal_polynomial} (top),~\ref{sec:treelevel_polynomial} (middle) 
and~\ref{sec:scaleinvariant} (bottom).}
\label{tab:frictioncoefficients}
\end{table}

Under the assumption that the transition strength satisfies $\alpha > \alpha_{\infty}$ (the case in which the bubbles undergo an accelerated expansion for part of the phase transition~\cite{Espinosa:2010hh,Bodeker:2009qy}), the $\gamma_{\rm eq}$ factor the bubble wall reaches when it ceases to accelerate is given by
\be
\gamma_{\rm eq}=\frac{\alpha-\alpha_\infty}{\alpha_{\rm eq}}\, .
\ee
The $\gamma$ factor that the wall would reach at the time of bubble collisions in the absence of plasma friction is simply expressed through the initial and final bubble radii as  
\be
\gamma_*=\frac{2}{3}\frac{R_*}{R_0}, \quad \, H\,R_*=v_w \left(8\pi\right)^{\frac{1}{3}}\left( \frac{\beta}{H}\right)^{-1}, \quad R_0=\left(\frac{3S_3}{2\pi \Delta V}\right)^{\frac{1}{3}} \, ,
\ee
where we have already expressed these conveniently through the known action and nucleation rate. The fraction of energy available to produce GWs from bubble collisions is then given by~\cite{Ellis:2019oqb}:
\be
\kappa_{\rm col}\approx\frac{3}{2}\frac{\gamma_{\rm eq}}{\gamma_*} \, .
\ee 

The energy fraction released by the phase transition that is converted into bulk fluid motion~\cite{Espinosa:2010hh} has already been given in~\eqref{eq:kappa_eff}, which for $\alpha > \alpha_{\infty}$ needs to be modified to subtract the energy going into acceleration of the wall instead of the plasma.    
In the limit of very relativistic walls $v_w\rightarrow 1$ (as expected for $\alpha > \alpha_{\infty}$), the relevant energy fraction for sound waves 
reads (see~\cite{Espinosa:2010hh,Ellis:2019oqb}) 
\be
\kappa_{\rm sw}|_{v_w\rightarrow 1} =\frac{\alpha_{\rm eff}}{\alpha} \frac{\alpha_{\rm eff}}{0.73+0.083\sqrt{\alpha_{\rm eff}}+\alpha_{\rm eff}} \,, \quad \alpha_{\rm eff} = \alpha(1-\kappa_{\rm col}) \,.
\ee
Finally, in order to obtain the GW spectrum from the phase transition as it would be observed today, we must include the corresponding redshift factors~\cite{Kamionkowski:1993fg}. The normalized energy density scales as:
\be
\begin{aligned} \label{eq:Omegaredshift}
\Omega_{{\rm GW},0} &= \left(\frac{a_*}{a_0}\right)^4  \left(\frac{H_*}{H_0}\right)^2 \Omega_{{\rm GW},*} 
=  1.67\times 10^{-5} h^{-2} \left(\frac{100}{g_{\rm eff}(T_{\rm reh})}\right)^\frac13 \Omega_{{\rm GW},*} \,,
\end{aligned}
\ee
while the frequency becomes:
\be
\begin{aligned} \label{eq:fredshift}
f_0 & = \frac{a_*}{a_0} f_* 
= 1.65\times 10^{-5} \,{\rm Hz}\, \left( \frac{T_{\rm reh}}{100\,{\rm GeV}} \right) \left( \frac{g_{\rm eff}(T_{\rm reh})}{100} \right)^\frac{1}{6} \left(\frac{f_*}{H_*}\right) \,,
\end{aligned}
\ee
where the star index denotes quantities at the end of the phase transition. The reheating temperature is approximately given by~\cite{Ellis:2018mja}
\be
T_n \left(1+\alpha \right)^\frac14 \, ,
\ee
and can be quite different from the temperature at which the phase transition begins, $T_n$, particularly for very strong transitions where a lot of energy is stored in the form of vacuum energy that later reheats the universe when the transition ends.  

\subsection{GW sources}

We now discuss the GW spectrum generated from sound waves and turbulence in the plasma, as well as from the collisions of scalar field bubbles. 
We start with the contribution to the GW signal from bubble collisions, as given approximately by~\cite{Cutting:2018tjt}:
\be
\Omega_{{\rm col},*} = 
0.3\, v_w^2 \,\left(\frac{\beta}{H}\right)^{-2}
\left(\frac{\kappa_{\rm col}\,\alpha}{1+\alpha}\right)^2 
\left(\frac{f_*}{f_{\rm col}}\right)^3\left[1+2\left(\frac{f_*}{f_{\rm col}}\right)^{2.07}\right]^{-2.18} \,,
\ee
with the peak frequency $f_{\rm col}$ given by (we suppress the star index in $\beta/H$ to match common notation)
\be
\frac{f_{\rm col}}{H_*} \simeq \frac{1.1}{v_w}\frac{\beta}{H}\, .
\ee
We turn next to GW sources related to the thermal plasma. Using the results from numerical lattice simulations~\cite{Hindmarsh:2013xza,Hindmarsh:2015qta,Hindmarsh:2017gnf}, the GW spectrum generated by sound waves propagating in the plasma after the transition may be approximated by~\cite{Caprini:2015zlo} (see also~\cite{Caprini:2019egz}):
\be \label{eq:Omegasw}
\Omega_{{\rm sw},*} = 
0.25\, v_w \, (H_*\tau_{\rm sw}) \left(\frac{\beta}{H}\right)^{-1}
\left(\frac{\kappa_{\rm sw} \,\alpha }{1+\alpha }\right)^2
\left(\frac{f_*}{f_{\rm sw}}\right)^3 \left[1+\frac{3}{4} \left(\frac{f_*}{f_{\rm sw}}\right)^2\right]^{-\frac72} \,,
\ee
with the peak frequency given by
\be
\frac{f_{\rm sw}}{H_*} \,\simeq\, \frac{1.16}{v_w}\frac{\beta}{H}
\, ,
\label{eq:frequencySW}
\ee
and the duration of the sound-wave period
\be
H_* \tau_{\rm sw} \equiv \min\left[1, H_* \tau_{\rm sh} \right]
\ee
with $H_* \tau_{\rm sh}$ given by~\eqref{eq:RHoverUf}.
%
%Details of the calculation of $U_f$ were discussed in Section~\ref{sec:hydrodynamics}. 
%
We nevertheless note that a recent analytical description of plasma sound waves~\cite{Hindmarsh:2019phv} yields somewhat different asymptotic behaviours when 
$f_* \gg f_{\rm sw}$ and $f_* \ll f_{\rm sw}$ from that in~\eqref{eq:Omegasw}.  
We also stress that the numerical lattice simulations of GW production from plasma sound waves have been carried out for small/moderate values of $\alpha$, 
and that the naive extrapolation to scenarios with large phase transition strengths used in this work 
(e.g.,~in the scale-invariant Coleman-Weinberg-like case) may not apply (see the discussion in~\cite{Caprini:2019egz}). These comments highlight the fact that the precise understanding of GW production from sound waves is still ongoing, and in that some care must be taken in interpreting our sound wave GW results. 

\vspace{2mm}

If the sound-wave period lasts much less than a Hubble time, $H_* \tau_{\rm sh} \ll 1$, there is a sizeable amount of energy in the bulk fluid motion when the flow becomes nonlinear. This energy could then (ideally) be available to create turbulence in the plasma, which also sources GWs.  
Assuming that all the energy in the bulk fluid motion remaining at the end of the sound wave period is transferred into turbulence, 
its contribution to the GW spectrum may be modelled as~\cite{Caprini:2009yp}:
\be \label{eq:Omegaturb}
\Omega_{{\rm turb},*} = 
19.9 \,  v_w \,
\left(\frac{\beta}{H}\right)^{-1} \left(1-H_*\tau_{\rm sw}\right) \left(\frac{\kappa_{\rm sw} \,\alpha }{1+\alpha }\right)^{3/2} 
\frac{\left(\frac{f_*}{f_{\rm turb}}\right)^3 \left[1+\left(\frac{f_*}{f_{\rm turb}}\right)\right]^{-\frac{11}{3}}}{1+8\pi f_*/H_*}\,,
\ee
with the peak frequency
\be
\frac{f_{\rm turb}}{H_*} \simeq \frac{1.75}{v_w} \left(\frac{\beta}{H}\right)
\,.
\ee
Given that~\eqref{eq:Omegaturb} assumes a perfect conversion of bulk fluid energy into turbulent energy, it may be viewed as an upper bound on the possible turbulent GW contribution, assuming a correct modelling of the turbulence as a GW source. 
In particular, this estimate neglects the possibility that some of that fluid energy is instead converted to plasma heat. It also does not take into account the recent studies on plasma vortical motion at the end of the phase transition~\cite{Cutting:2019zws}. These indicate that for slow deflagrations with sizable $\alpha$, plasma turbulence can be efficiently generated together with sound waves during the linear evolution of the fluid, and also suggest that 
the conversion efficiency of vortical energy into GWs is significantly smaller than that of sound waves. All these aspects should be kept in mind when interpreting our results for GWs from turbulence.

\subsection{Parameter space scans and results}

We have performed scans over the parameter spaces of the three scenarios discussed in Section~\ref{sec:genral_polynomial}, using the ranges defined there for each of them. 
The scans have been done separately for three proposed experiments: LISA~\cite{Audley:2017drz,Bartolo:2016ami}, the Einstein Telescope~(ET)~\cite{Punturo:2010zz,Hild:2010id} and AEDGE~\cite{Bertoldi:2019tck}, using in each case a Monte Carlo Markov chain aimed at maximizing the signal-to-noise ratio (SNR) for a given GW experiment, by varying the parameters a hundred times around initial scan point values.

\begin{figure}
\begin{center}
\includegraphics[width=0.49\textwidth]{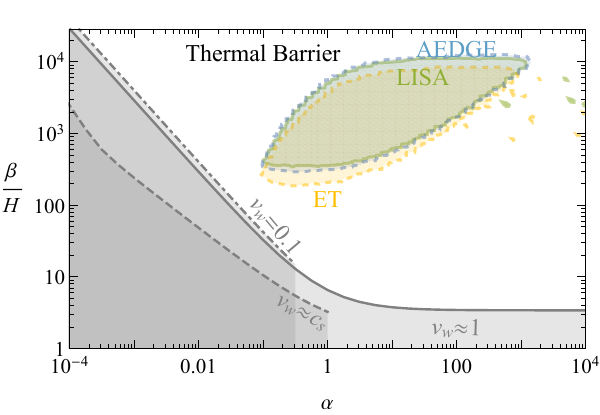}
\includegraphics[width=0.49\textwidth]{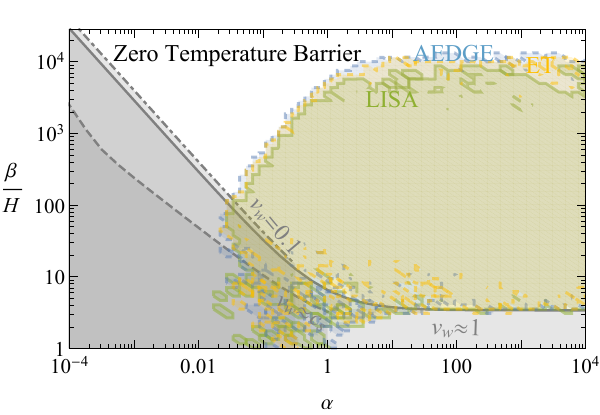}
\includegraphics[width=0.49\textwidth]{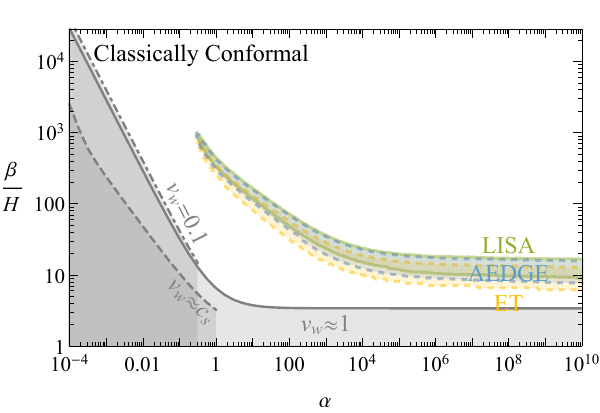}
\end{center}
\vspace{-5mm}
\caption{\it The coloured shading shows the regions of the $(\alpha, \beta/H)$ plane featuring a $SNR>10$ for  LISA~\cite{Audley:2017drz,Bartolo:2016ami} (green), the Einstein Telescope~(ET)~\cite{Punturo:2010zz,Hild:2010id} (orange) and AEDGE~\cite{Bertoldi:2019tck} (blue), for potential with a pure thermal barrier (upper left), a zero-temperature barrier (upper right) and a scale-invariant Coleman-Weinberg scenario (lower). The grey shading shows the regions with long-lived GW emission from sound waves ($H \tau_{\rm sh} > 1$) for different bubble wall velocities.
}
\label{fig:SNRGWscan}
\end{figure}

The results of these scans are shown in Fig.~\ref{fig:SNRGWscan} as coloured areas in the $(\alpha, \beta/H)$ planes satisfying $\mathrm{SNR} \geq 10$ for LISA (green), ET (orange) and AEDGE (blue), respectively, for the scalar potential with a pure thermal barrier from Section~\ref{sec:thermal_polynomial} (upper left), the scalar potential with a zero-temperature barrier from Section~\ref{sec:treelevel_polynomial} (upper right) and the scale-invariant Coleman-Weinberg potential from Section~\ref{sec:scaleinvariant} (lower). In all the scans we have assumed
$v_w \sim 1$ when computing the resulting GW signal, which corresponds to the most optimistic choice for GW production.~\footnote{We note that Fig.~\ref{fig:SNRGWscan} shows the regions where sound waves are long-lasting ($H \tau_{\rm sh} > 1$) for various values of $v_w$. Whilst smaller values of $v_w$ increase the $\beta/H$ range of long-lasting sound waves (for a given $\alpha$), decreasing $v_w$ would also significantly diminish the amplitude of the resulting GW signal.}
We stress that, despite Fig.~\ref{fig:SNRGWscan}  giving the naive impression that the sensitivity ranges of the GW experiments LISA, ET and AEDGE are fairly similar, this is the casee only for $\beta/H$ and $\alpha $, but not for the temperature of the transition $T_*$, which determines the characteristic frequency of the GW signal.

As shown in the upper left panel of Fig.~\ref{fig:SNRGWscan}, the case of a purely thermal barrier does not allow sound waves to last a significant fraction of a Hubble time. In contrast, for the scenario with a zero-temperature barrier (upper right panel) this is possible for low values of $\beta/H$ and moderate to large $\alpha$. It is also important to note that for such supercooled transitions we always have $\alpha>\alpha_{\infty}$, so that $v_w \to 1$ is indeed expected in those cases. Finally, in the lower panel of Fig.~\ref{fig:SNRGWscan} we see that in scale-invariant Coleman-Weinberg-like models there is only a restricted band of the $(\alpha, \beta/H)$ plane where GW signals with SNR $> 10$ can be achieved which, as already shown in Section~\ref{sec:scaleinvariant}, lies well within the region where sound waves only last for a small fraction of a Hubble time.

\begin{figure}[h]
\begin{center}
\includegraphics[width=0.49\textwidth]{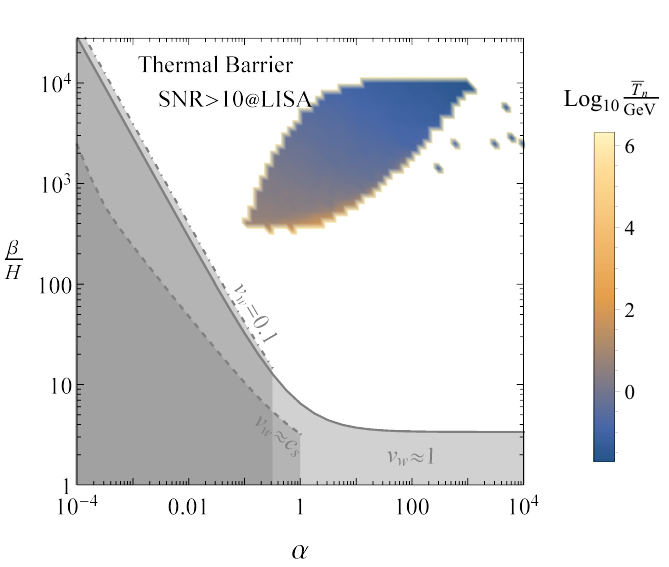}
\includegraphics[width=0.49\textwidth]{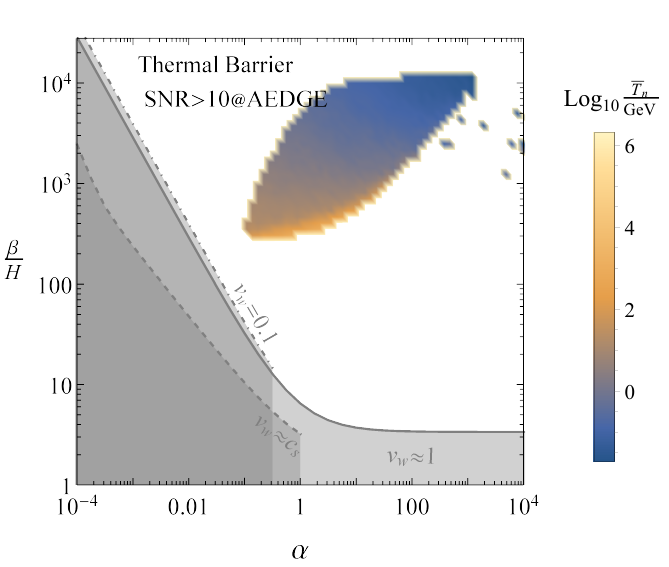}
\includegraphics[width=0.49\textwidth]{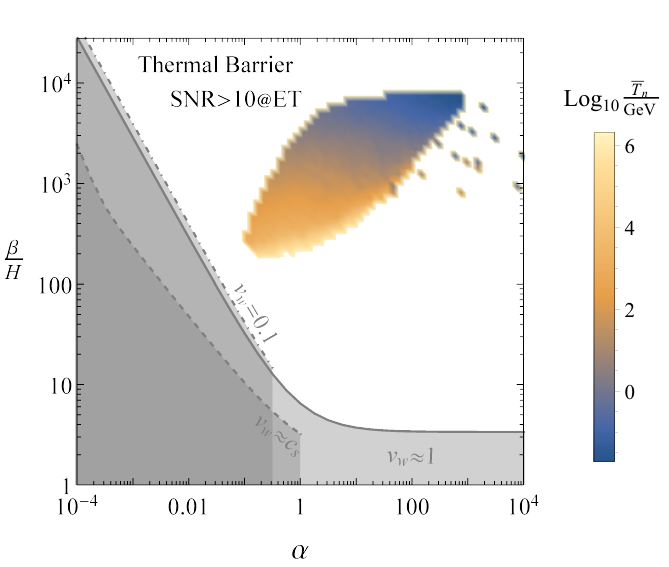}
\end{center}
\vspace{-5mm}
\caption{\it The coloured regions of the $(\alpha, \beta/H)$ planes feature a $SNR>10$ for  LISA (upper left panel), AEDGE (upper right panel) and ET (lower panel) for a potential with a pure thermal barrier. The color gradient corresponds to different values of the (average) phase transition nucleation temperature $T_n$ in each case. The grey shading shows the regions with long-lived GW emission from sound waves ($H \tau_{\rm sh} > 1$) for different bubble wall velocities.
}
\label{fig:Tnplots}
\end{figure}

As already mentioned, whilst the capabilities of LISA, ET and AEDGE to detect a GW signal from a  first-order cosmological phase transition appear similar for $\alpha$ and $\beta/H$, they are sensitive to very different transition temperatures.
%
%Judging only by Fig.~\ref{fig:SNRGWscan} one might get the impression that the capabilities of the different experiments are quite similar. This is, however, an artefact of the choice of parameters on the axes, as in practice the regions within reach of each experiment in this figure correspond to very different values of the parameters in the scalar potentials.
% 
To illustrate this explicitly, we show in Fig.~\ref{fig:Tnplots} the average nucleation temperature $T_n$ probed by the LISA (upper left panel), AEDGE (upper right panel) and ET (bottom panel) GW experiments, for the scenario with a purely thermal barrier discussed in Sec.~\ref{sec:thermal_polynomial}, as a function of $\alpha$ and $\beta/H$, clearly showing the expected correlation, with experiments sensitive to lower frequency ranges being more sensitive to lower transition temperatures. 

\vspace{2mm}

\begin{figure}
\begin{center}
\includegraphics[width=0.75\textwidth]{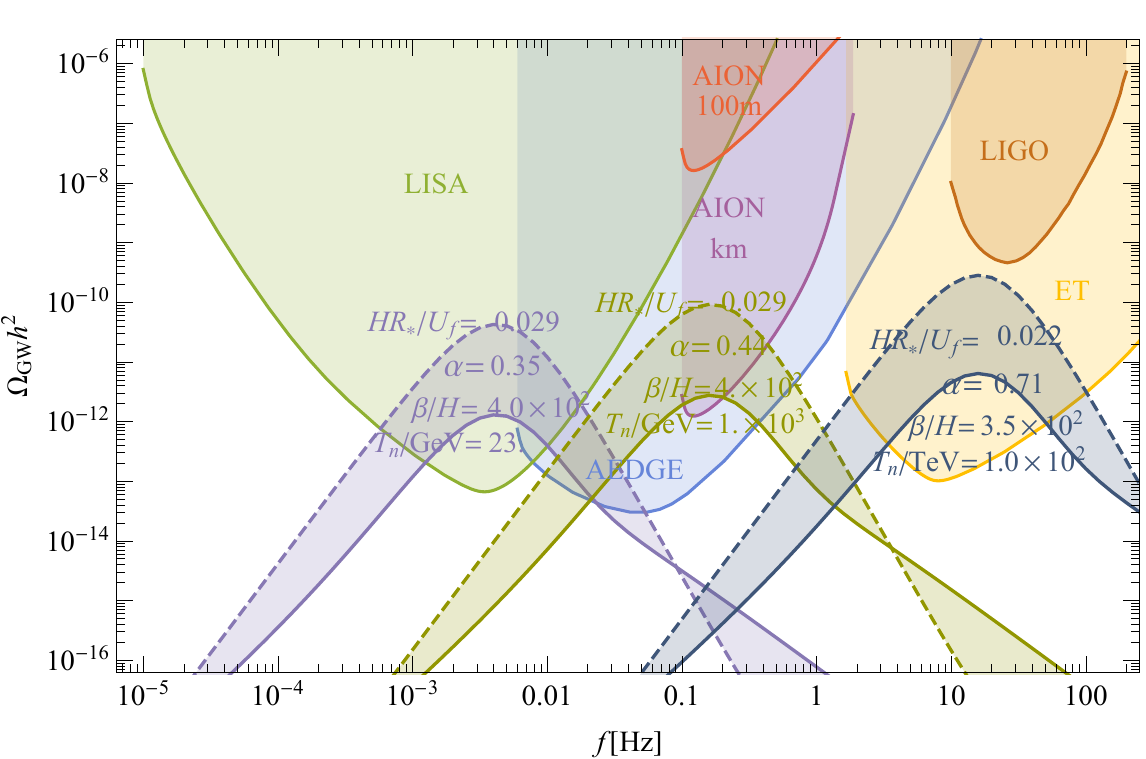}
\end{center}
\vspace{-5mm}
\caption{\it GW spectra for parameter choices with some of the highest SNR for LISA~\cite{Audley:2017drz,Bartolo:2016ami}, the Einstein Telescope~(ET)~\cite{Punturo:2010zz,Hild:2010id} and AEDGE~\cite{Bertoldi:2019tck} found in our scan of the scenario with a purely thermal potential barrier from Sec.~\ref{sec:thermal_polynomial}. Solid lines show GW spectra as calculated in this work, while dashed lines show GW spectra calculated using older estimates from~\cite{Caprini:2015zlo}, which assumed that the sound wave period lasts for a Hubble time, i.e., $H \tau_{\rm sh} > 1$. We also show the sensitivities of LIGO~\cite{TheLIGOScientific:2014jea,Thrane:2013oya,TheLIGOScientific:2016wyq} and the terrestrial atom interferometer AION~\cite{Badurina:2019hst} (which is similar to that of MAGIS~\cite{Graham:2016plp,Graham:2017pmn}).
}
\label{fig:SWreductionplot1}
\end{figure}

\begin{figure}
\begin{center}
\includegraphics[width=0.75\textwidth]{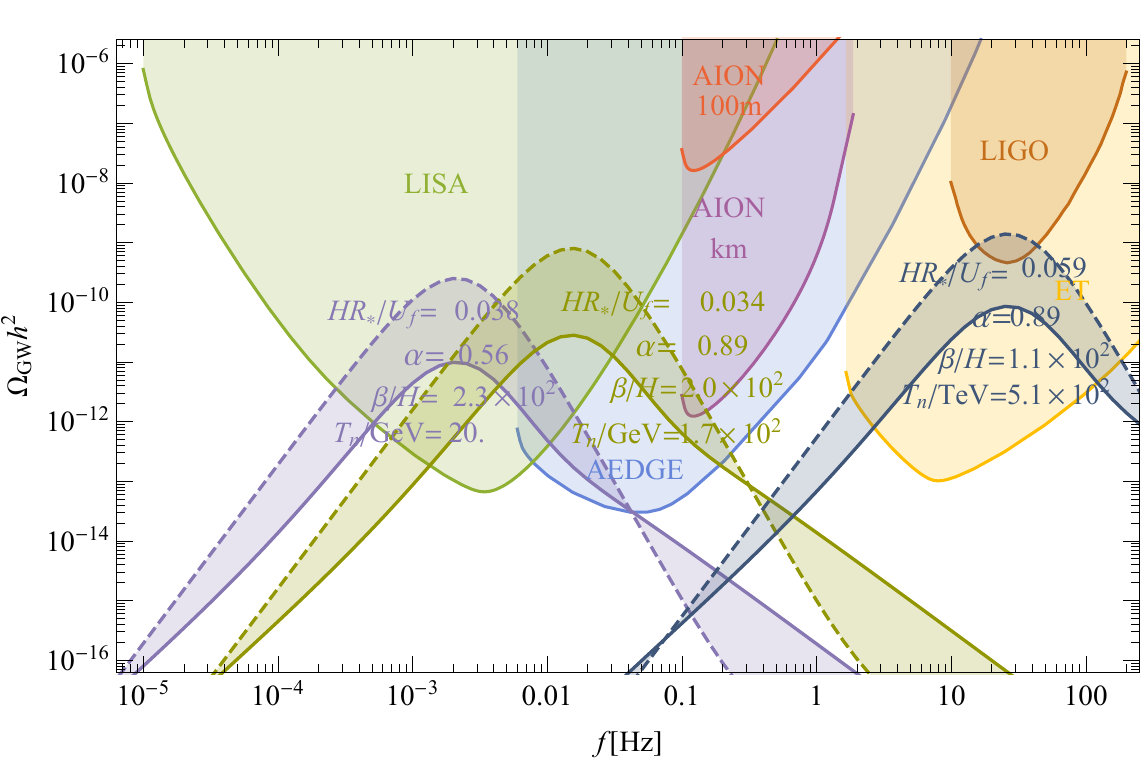}
\end{center}
\vspace{-5mm}
\caption{\it Same as Fig.~\ref{fig:SWreductionplot1}, for the scenario with a zero-temperature barrier from Section~\ref{sec:treelevel_polynomial}.
}
\label{fig:SWreductionplot2}
\end{figure}

\begin{figure}
\begin{center}
\includegraphics[width=0.75\textwidth]{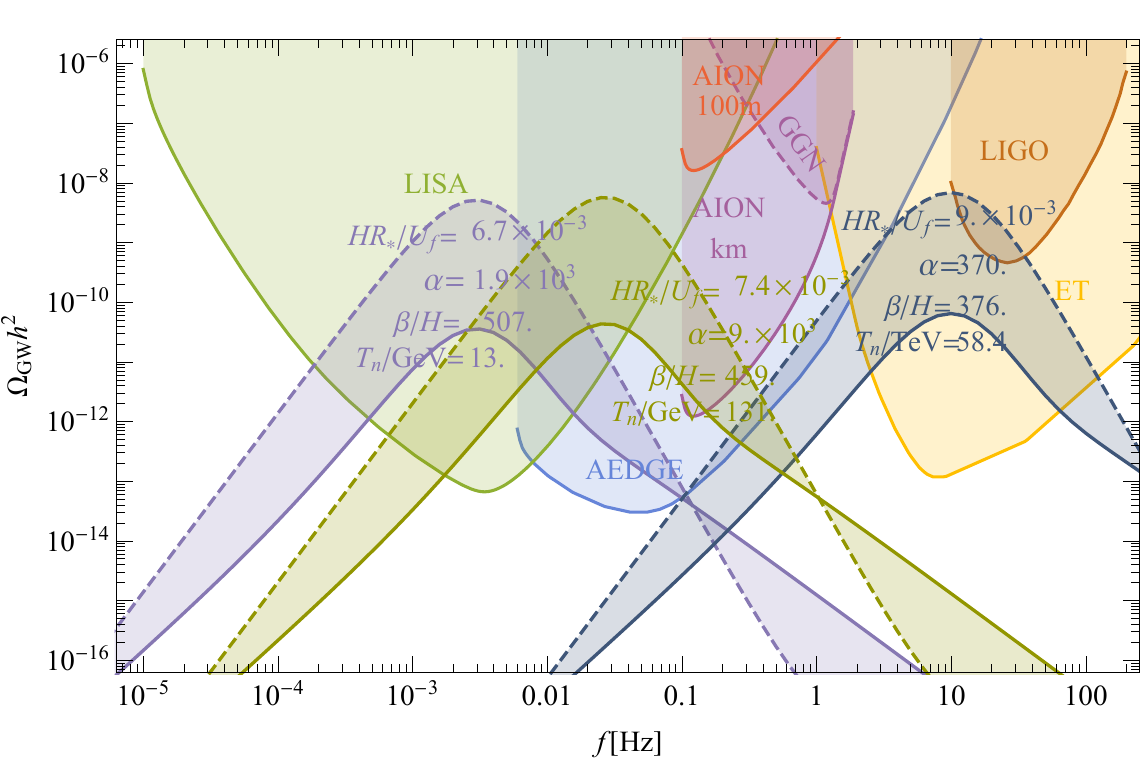}
\end{center}
\vspace{-5mm}
\caption{\it Same as Fig.~\ref{fig:SWreductionplot1}, for the scenario with a scale-invariant Coleman-Weinberg-like potential from Section~\ref{sec:scaleinvariant}.
}
\label{fig:SWreductionplot3}
\end{figure}

Finally, we also show for illustration in Figs.~\ref{fig:SWreductionplot1}, \ref{fig:SWreductionplot2} and \ref{fig:SWreductionplot3} several GW spectra (solid lines) that would yield a high SNR in the LISA, AEDGE and ET experiments
(we also depict the sensitivities of LIGO~\cite{TheLIGOScientific:2014jea,Thrane:2013oya,TheLIGOScientific:2016wyq} and the proposed terrestrial atom interferometer AION~\cite{Badurina:2019hst}, which has a similar sensitivity to that of the MAGIS~\cite{Graham:2016plp,Graham:2017pmn} experiment)\footnote{We show the standard power-law integrated sensitivity~\cite{Thrane:2013oya}, and refer the reader to~\cite{Schmitz:2020syl,Alanne:2019bsm} for newer alternatives focusing specifically on phase transitions (see also the analysis from~\cite{Caprini:2019pxz}).}
in each of the models we have studied respectively in Sections~\ref{sec:thermal_polynomial}, \ref{sec:treelevel_polynomial}, and~\ref{sec:scaleinvariant}. In each case, we show for 
comparison (dashed lines) the GW spectra that would be calculated assuming that the sound waves could act as a GW source during an entire Hubble time. We see that in all cases the possible GW signals are reduced by one to two orders of magnitude when a more realistic estimate of the duration of the sound wave GW source is used.
Even with these reductions in GW production due to the curtailment of the period of sound waves as an active GW source, we see that each of the three scenarios may produce detectable signals in any of LISA, AEDGE and ET. We note also that for some model parameters there are significant opportunities for measurements in a pair of detectors, illustrated by the facts that the optimal parameter choices for LISA would also yield observable signals in AEDGE, and vice versa, opening up the possibility of measuring the GW spectrum over a wide range of frequencies (see, e.g.,~\cite{Figueroa:2018xtu} for a discussion). However, we note that the reductions in the GW strength diminish significantly the possibilities for MAGIS/AION and particularly for LIGO to detect GWs from a phase transition in the early Universe.

%%%%%%%%%%%%%%%%%%%%%%%%%%%%%%%%%%%%%%%%%%%%%%%%%%%%%%%%%%%%%%%%%%%%%%%%%%%%%%%%%%%%%%%%%%%%%%%%%%%%%%%%%
%%%%%%%%%%%%%%%%%%%%%%%%%%%%%%%%%%%%%%%%%%%%%%%%%%%%%%%%%%%%%%%%%%%%%%%%%%%%%%%%%%%%%%%%%%%%%%%%%%%%%%%%%
%%%%%%%%%%%%%%%%%%%%%%%%%%%%%%%%%%%%%%%%%%%%%%%%%%%%%%%%%%%%%%%%%%%%%%%%%%%%%%%%%%%%%%%%%%%%%%%%%%%%%%%%%
%%%%%%%%%%%%%%%%%%%%%%%%%%%%%%%%%%%%%%%%%%%%%%%%%%%%%%%%%%%%%%%%%%%%%%%%%%%%%%%%%%%%%%%%%%%%%%%%%%%%%%%%%
%%%%%%%%%%%%%%%%%%%%%%%%%%%%%%%%%%%%%%%%%%%%%%%%%%%%%%%%%%%%%%%%%%%%%%%%%%%%%%%%%%%%%%%%%%%%%%%%%%%%%%%%%

\section{Conclusions}
\label{sec:conx}

We have analyzed in this work the possible duration of a sound wave GW signal from a cosmological first-order phase transition, using simple polynomial forms of effective potential without and with a potential barrier at zero temperature, as well as a scale-invariant Coleman-Weinberg-like model. We have taken into account the various possible hydrodynamical solutions for the expansion of bubbles during the transition, namely deflagrations, detonations and hybrid solutions, depending on the value of the bubble expansion velocity $v_w$, obtaining in each case the energy available for GW production in terms of $v_w$ and the strength of the transition $\alpha$. We have considered the emission of GWs by bubble collisions as well as by sound waves and by turbulence in the plasma following the transition, assuming for the latter that all the energy in the bulk fluid motion when the flow becomes nonlinear is available to create turbulence. Scanning over the parameter space of the different scenarios and imposing systematically the requirement of percolation, we then identified the regions of the $(\alpha, \beta/H)$ plane (where $\beta$ is the time scale for the transition) in which the future GW experiments LISA, the Einstein Telescope~(ET) and AEDGE could detect GWs with a large signal-to-noise ratio, $\mathrm{SNR} > 10$.
The complementarity among the approved GW LISA experiment, the planned ET experiment and the proposed AEDGE atom interferometer space experiment is manifest in our Figs.~\ref{fig:SWreductionplot1}, \ref{fig:SWreductionplot2} and \ref{fig:SWreductionplot3}. LISA has optimal sensitivity at frequencies $\lesssim 10^{-2}$~Hz, whereas ET and AEDGE would have optimal sensitivities at frequencies $\sim 10$~Hz and frequencies $\sim 10^{-1}$~Hz, respectively. A combination of them might be needed to map out the full GW spectrum in any given model, as highlighted in our Figs.~\ref{fig:SWreductionplot1}, \ref{fig:SWreductionplot2} and \ref{fig:SWreductionplot3}. 
As seen in these Figures, the proposed 1-km versions of the terrestrial atom interferometers MAGIS and/or AION could also have some sensitivity to the models discussed in this paper.

\vspace{1mm}

When the first-order phase transition is of purely thermal origin (i.e.,~there is no zero-temperature barrier in the scalar potential), the regions of parameter space with $\mathrm{SNR} > 10$ in the different experiments all correspond to a duration of the GW production from sound waves that is significantly less than a Hubble time. This is also true for most of the $(\alpha, \beta/H)$ parameter space of models with a zero-temperature potential barrier, although in this case there are small regions of parameter space where the duration may extend up to a Hubble time. 
%We find that only less than about $3\times 10^{-4}$ of all the points in our scans fall in this region for any of the experiments. 
All of the points in this region also correspond to substantial phase transition supercooling. %, with $T_* < 10$ GeV.
The GW spectra calculated in this work in are generally of smaller amplitude than those calculated using results from~\cite{Caprini:2015zlo}, which assumed that the sound-wave period lasts a Hubble time, i.e., $H \tau_{\rm sh} = 1$. On the other hand, our GW spectra have broader extensions to higher frequencies, sourced by turbulence.

We stress that we have endeavoured to make our analysis as general as possible, extending the analysis of~\cite{Ellis:2018mja,Caprini:2019egz} in the context of particular particle physics models. We have thus covered generic models whose effective potentials can be approximated by a polynomial as well as classically conformal, Coleman-Weinberg-like models. Although all these models generally predict short bursts of GWs from sound waves lasting much less than a Hubble time, we recognise the phenomenological interest of other forms of potential to which our analysis does not apply directly. Other shortcomings of our analysis include our assumption that all the energy in the bulk fluid motion when the flow becomes nonlinear is available to create turbulence that can source GWs, and uncertainties in the modelling of GW production from turbulence. Since the GW production during the nonlinear evolution of the plasma at the end of the phase transition is not yet known, our assumptions may not hold in general.
These caveats aside, our analysis provides some encouragement that, despite the shortened sound wave period we find, future GW experiments may be able to probe the dynamics of a phase transition in the early universe over a wide range of frequencies, giving us direct access to the very early stages of the evolution of the Universe. 
%, opening a window on particle models that complements collider and other experiments.

\acknowledgments
The work of JE and ML was supported by the UK STFC Grant ST/P000258/1. JE was also supported by the Estonian Research Council via a Mobilitas Pluss grant and ML by Polish National Science Center grant 2018/31/D/ST2/02048. J.M.N. was supported by Ram\'on y Cajal Fellowship contract RYC-2017-22986, and also acknowledges support from the Spanish MINECO's ``Centro de Excelencia Severo Ochoa" Programme under grant SEV-2016-0597, from the European Union's Horizon 2020 research and innovation programme under the Marie Sklodowska-Curie grant agreements 690575  (RISE InvisiblesPlus) and 
674896 (ITN ELUSIVES) and from the Spanish Proyectos de I$+$D de Generaci\'on de Conocimiento via grant PGC2018-096646-A-I00.

\appendix

\section{Further aspects of hydrodynamics}
\label{sec:Hydro_Appendix}

We identified in the main text three possibilities for the hydrodynamics of the transition, depending upon the bubble wall velocity $v_w$, which we consider here in turn.

\begin{itemize}
\item For wall velocities slower than the speed of sound of the plasma ($v_w < c_s$) we obtain a {\it deflagration}, in which the wall simply pushes the plasma shell around it;

\item For larger wall velocities $v_w>c_s$, if the strength of the transition is sufficiently weak we have a {\it detonation} solution, in which the plasma shell follows the expanding wall;

\item Finally for $v_w>c_s$ and a strong interaction with the plasma, we find a {\it hybrid} solution that, as the name suggests, combines the two behaviours mentioned above.   
\end{itemize}

\subsection{Detonations}

We consider first the simplest case of the detonation solution. In this case the plasma shell follows the bubble wall, so that in the wall frame the plasma outside simply goes into the wall with speed $v_+=v_w\,$. We can then find the velocity inside using Eq.~\eqref{eq:vfluid}, which yields
\begin{equation}
    v_-=\frac{1-3\alpha+3 v_w^2\left(1+\alpha \right)+\sqrt{\left(1-3\alpha+3 v_w^2\left(1+\alpha \right)\right)^2-12 v_w^2} }{6 v_w}\, ,
\end{equation}
where we have set $\alpha_+=\alpha$ (see Eq.~\eqref{eq:alpha}), as in this solution the bubble does not perturb the plasma outside and the value in our matching condition is the same as the global strength of the transition in the symmetric phase. For the above equation to have a positive solution we need
\begin{equation} \label{eq:alpha_max_deto}
    \alpha<\alpha_{\rm max}^{\rm det}=\frac{\left(1-\sqrt{3}v_w\right)^2}{3\left(1-v^2_w\right)}\, .
\end{equation}
If this is not fulfilled, a detonation solution is not realised, and we find instead a hybrid solution, which we discuss in detail in the last Section of this Appendix.

\begin{figure}[h]
\begin{center}
\includegraphics[width=0.98\textwidth]{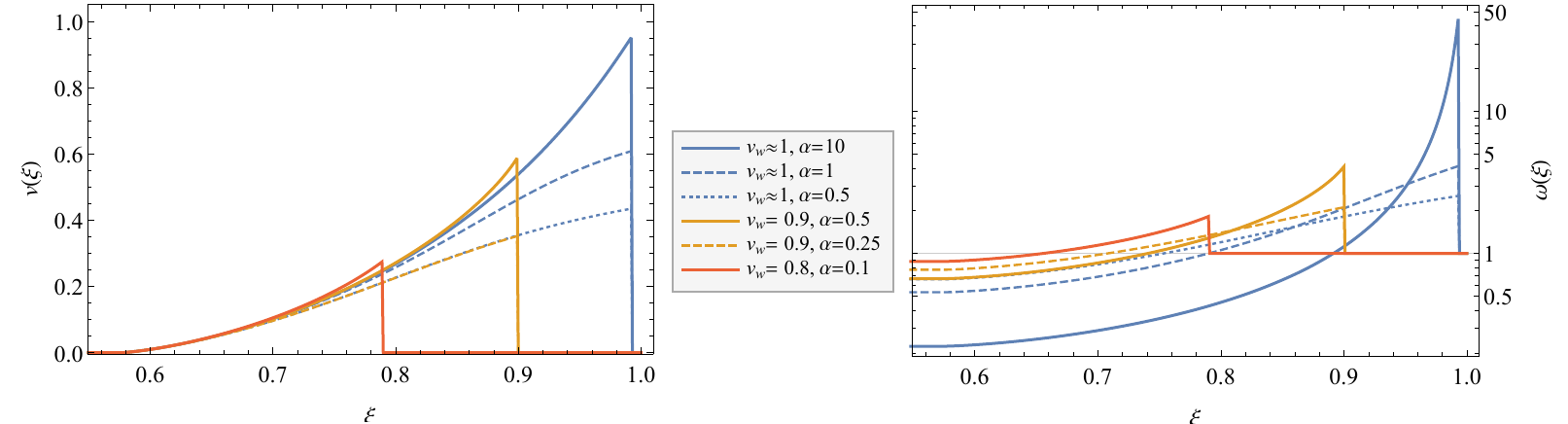}
\end{center}
\vspace{-5mm}
\caption{\it Left panel: Detonation plasma velocity profiles for three wall velocities and several possible strengths of the transition. 
Right panel: Enthalpy profile examples, normalised to the enthalpy in the symmetric background phase.}
\label{fig:detprofiles}
\end{figure}

We are now ready to find the velocity and enthalpy profiles in the frame of the bubble. We simply need to solve Eqs.~\eqref{eq:vfluid} and~\eqref{eq:enthalpy} inside the bubble, starting from the plasma velocity at the wall transformed into the bubble frame, and using the enthalpy given by the matching conditions:    
\begin{equation}
    v(\xi<v_w)=\frac{v_w-v_-}{1-v_w v_-}\, ,
    \quad \quad 
     \omega(\xi<v_w)=\frac{v_w}{1-v_w^2}\frac{1- v_-^2}{v_-}
    \, .
\end{equation}
The velocity goes smoothly to zero at $\xi=v_-$, and from that point remains constant just as $\omega$ as we approach the centre of the bubble ($\xi\rightarrow 0$).
We show examples of resulting profiles in Fig.~\ref{fig:detprofiles}.

%%%%%%%%%%%%%%%%%%%%%%%%%%%%%%%%%%%%%%%%%%%%%%%%%%%%%%%%%%%%%%%%%%%%%%%%%%%%%%%
\subsection{Deflagrations}

The criterion for the realisation of a deflagration solution is simply that the wall velocity be subsonic: $\xi_{w}<c_s\,$. The plasma is at rest inside the wall in the frame of the bubble centre, which means that in the wall frame $v_-=\xi_w$, while the velocity outside is found using Eq.~\eqref{eq:vfluid}. The immediate issue here is that $\alpha_+ \neq \alpha$, as the latter is defined in the symmetric phase background and not in the heated plasma shell. Thus, we start with some value for $\alpha_+$ and find $\alpha$ as an output of our calculation. We can then find the profile for a given $\alpha$ by scanning numerically over our input $\alpha_+$ to find the desired strength of the transition. 

\begin{figure}[h]
\begin{center}
\includegraphics[width=0.98\textwidth]{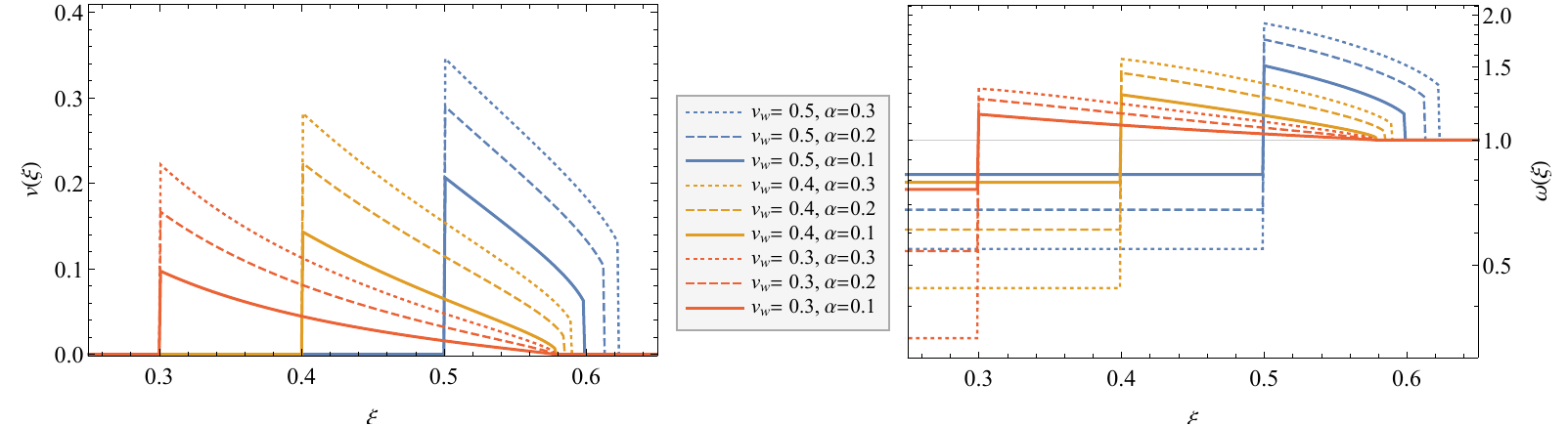}
\end{center}
\vspace{-5mm}
\caption{\it Left panel: Deflagration plasma velocity profiles for three wall velocities and several possible strengths of the transition. 
Right panel: Enthalpy profile examples, normalised to the enthalpy in the symmetric background phase.}
\label{fig:defprofiles}
\end{figure}

Next we need to solve Eqs.~\eqref{eq:vfluid} outside the wall, starting with
\begin{equation}
    v(\xi>v_w)=\frac{v_w-v_+}{1-v_w v_+}
    \, .
\end{equation}
The extra difficulty is that the only consistent solution ends with a shock front rather than the velocity decreasing smoothly. The shock front appears at $\xi_{\rm sh}$ defined by the condition
\be \label{eq:xi_shock}
0=\left. \frac{c_s^2-\xi^2}{(c_s^2-1)\xi}-v(\xi)\right|_{\xi=\xi_{\rm sh}} \, .
\ee
The jump in velocity means that we must use again the matching conditions~\eqref{eq:matching_conditions}, this time for inside and outside the shock front.
This also leads to an extra jump in the enthalpy profile, which we again solve inside the shock front using Eq.~\eqref{eq:enthalpy}, and starting from
\be \label{eq:omega_sh}
\omega(\xi_{\rm sh})=\frac{\xi_{\rm sh}}{1-\xi_{\rm sh}^2}\frac{1-v_{\rm sym}^2}{v_{\rm sym}} \, , \quad v_{\rm sym}=\frac{\xi_{\rm sh}-v(\xi_{\rm sh})}{1-\xi_{\rm sh} v(\xi_{\rm sh})} \, ,
\ee
where the second equality gives the velocity outside the shock front in the symmetric phase background.
We can now use this result to match to the outside of the plasma shell and derive the strength of the transition to which our inputs correspond:
\be
\alpha=\alpha_+ \, \omega(\xi=v_w)\, .
\ee
To calculate the enthalpy inside the bubble we need to include also the step at the bubble wall:
\be
\omega(\xi<v_w)=\omega(\xi>v_w) \, \frac{1-v_w^2}{v_w} \, \frac{v_+}{1-v_+^2}\, ,
\ee
after which the enthalpy remains constant all the way to the centre of the bubble. 
We show examples of results from this calculation in the forms of velocity and enthalpy profiles in Fig.~\ref{fig:defprofiles}.

\subsection{Hybrids}

We describe finally hybrid solutions that combine the features discussed in the previous subsections. These solutions are realised for supersonic wall velocities $v_w>c_s$ if the transition is so strong that the bound~\eqref{eq:alpha_max_deto} is violated.
This would mean we are dealing with a supersonic deflagration, as was shown in hydrodynamic simulations to be unstable and indeed to develop a rarefaction wave identical to a detonation~\cite{KurkiSuonio:1995pp}.

As before, we start with the velocity profile computed using~\eqref{eq:vfluid}.
Matching the velocities on both sides with detonation and deflagration constraints we see the wall velocity cannot be identified with the plasma velocity on either side. The only consistent choice is to set $v_-=c_s$ with $v_+$ given by~\eqref{eq:matching_wall}. Just as in the deflagration case, $\alpha_{out} \neq \alpha$ due to the appearance of a shock front. Thus we again have to use $\alpha_{out}$ as our input and proceed to calculate the corresponding $\alpha$.

To find the velocity inside the wall we start at the wall with
\begin{equation}
    v(\xi<v_w)=\frac{v_w-v_-}{1-v_w v_-}
    \, , \quad
      v(\xi>v_w)=\frac{v_w-v_+}{1-v_w v_+}
      \, .
\end{equation}
We solve eq.~\eqref{eq:vfluid} for $\xi=c_s$ inside the wall where the velocity reaches zero, and outside the bubble we solve until reaching the shock front, again defined by Eq.~\eqref{eq:xi_shock}, at which the velocity drops to zero.

To find the enthalpy profile normalised to the symmetric phase outside the shock front, we use Eq.~\eqref{eq:omega_sh} just as in the deflagration case. We then again solve Eq.~\eqref{eq:enthalpy} until the bubble wall, where we introduce the necessary step
\be
\omega(\xi<v_w)=\frac{1-v_-^2}{v_-} \, \frac{v_+}{1-v_+^2} \, \omega(\xi>v_w)\, ,
\ee
and then continue to $\xi=c_s$, beyond which enthalpy does not change any more going towards the centre of the bubble. 

\begin{figure}[h]
\begin{center}
\includegraphics[width=0.98\textwidth]{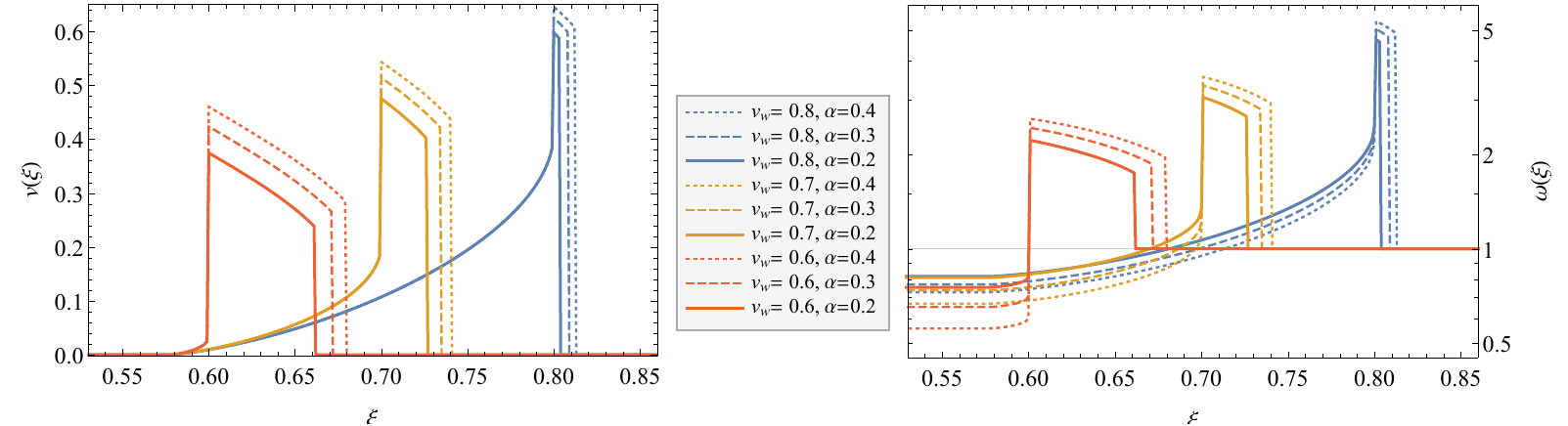}
\end{center}
\vspace{-5mm}
\caption{\it Left panel: Hybrid plasma velocity profiles for three wall velocities and several possible strengths of the transition. 
Right panel: Enthalpy profile examples, normalised to the enthalpy in the symmetric background phase.}
\label{fig:hybprofiles}
\end{figure}

As in the deflagration case, we find the strength of the transition corresponding to our input $\alpha=\alpha_+\omega(\xi>v_w)$. We then find numerically the inputs matching the desired strength of the transition.
Examples of the resulting velocity and enthalpy profiles are shown in Fig.~\ref{fig:hybprofiles}.

%%%%%%%%%%%%%%%%%%%%%%%%%%%%%%%%%%%%%%%%%%%%%%%%%%%%%%%%%%%%%%%%%%%%%%%%%%%%%
%%%%%%%%%%%%%%%%%%%%%%%%%%%%%%%%%%%%%%%%%%%%%%%%%%%%%%%%%%%%%%%%%%%%%%%%%%%%%
%%%%%%%%%%%%%%%%%%%%%%%%%%%%%%%%%%%%%%%%%%%%%%%%%%%%%%%%%%%%%%%%%%%%%%%%%%%%%
\section{Percolation and successful completion of the transition} 
\label{sec:percolation}

In specific models we encounter possible transitions with significant supercooling.
In these cases the simple exponential nucleation approximation $\Gamma \propto e^{-\beta t}$ can fail (see eq. \eqref{eq:beta_Taylor}), which is typically signalled by low values of $\beta/H$.
In order to avoid this problem in our analysis we use the more refined analysis of percolation described in detail in~\cite{Ellis:2018mja}.   

%%%%%%%%%%%%%%%%%%%%%%%%%%%%%%%%%%%%%%%%%%%%%%%%%%%%%%%%%%%%%%%%%%%%%%%%%%%%%
Assuming we know the decay rate $\Gamma$,
the probability of a given point still being in the false vacuum reads~\cite{Guth:1979bh,Guth:1981uk}
\be
\label{eq:prob_false_vacuum}
P(t)=e^{-I(t)}, \quad
I(t)=\frac{4\pi}{3}\int_{t_c}^{t} dt' \,\Gamma(t')\, a(t')^3\, r(t,t')^3\,,
\ee
where $a$ is the FRW scale factor and $r(t,t')$ is the radius of a bubble  
that nucleated at $t'$ after growing until time $t$:
\be \label{eq:bubblesizedef}
r(t,t') = \int_{t'}^{t}\frac{v_w\,d\tilde{t}}{a(\tilde{t}\hspace{0.4mm})} \,.
\ee
To solve for the scale factor we need the Friedmann equation including the radiation energy density $\rho_{\rm R}$ (we assume instantaneous reheating)  and the energy density of the unstable vacuum $\rho_{\rm V}$:
\be \label{eq:Hubble}
H(T)^2=\frac{1}{3 M_{\mathrm{pl}}^2}\left(\rho_{\rm R}+\rho_{\rm V} \right) = \frac{1}{3 M_{\mathrm{pl}}^2} \left(\frac{T^4}{\xi_g^2} + \Delta V \right) 
\,,
\ee
with $\xi_g=\sqrt{30/(\pi^2 g_*)}$, where $g_*=106.75$ is the number of degrees of freedom, and $M_{\mathrm{pl}} = 2.435 \times 10^{18}$ GeV. 
Using Equation~\eqref{eq:prob_false_vacuum}, the integrand of Equation~\eqref{eq:bubblesizedef} can be conveniently rewritten as function of temperature in terms of the product
\be   \label{eq:RV}
a(T') \, r(T,T') = a(T')\,\int_{T}^{T'}\frac{v_w\, d\tilde{T}}{\tilde{T}\,H(\tilde{T})\, a(\tilde{T})}
= \frac{1}{T'}\,\int_{T}^{T'}\frac{v_w\,d\tilde{T}}{H(\tilde{T})} \, ,
\ee
which can readily be computed in terms of elliptic functions, yielding
\be \label{eq:prob_false_vacuum_2}
I(T)  = \frac{4\pi}{3} \int^{T_c}_{T} 
\frac{dT'\,\Gamma(T')}{T'^4\,H(\tilde{T})}\,\left(\int_{T}^{T'}\frac{v_w\,d\tilde{T}}{H(\tilde{T})} 
\right)^3 \, .
\ee
We can now define the time of percolation as $P(T_*)=1/e$, corresponding to $I(T_*)=1$~\cite{Enqvist:1991xw}.
Finally, the necessary requirement for successful completion of the transition concerns the physical 
volume of the false vacuum $\mathcal{V}_{\rm false}\propto a(t)^3P(t)$, which starts decreasing at the percolation~\cite{Turner:1992tz}.
This is a strong requirement, particularly in cases where vacuum energy dominates the expansion, since the probability $P(t)$ now has to decrease faster than 
the increase in the volume of the inflating space. This condition can be conveniently rewritten as
\be \label{eq:falsevacuumvol}
\left. \frac{1}{\mathcal{V}_{\rm false}}\frac{d \mathcal{V}_{\rm false}}{d t}\right|_{t=t_*}
=\left. 3 H(t)- \frac{d I(t)}{d t}\right|_{t=t_*}=\left. H(T)\left( 3+T\,\frac{d I(T)}{dT} \right)\right|_{T=T_*} < 0 \, .
\ee
In our analysis of concrete models in the remainder of this Section we always calculate the percolation temperature and check that the above constraint is fulfilled.   

Let us also point out that in cases with a lot of supercooling the simple relation between the time sale and bubble size $H\, R_* =(8\pi)^\frac13 (\beta/H)^{-1}$ breaks down together with the definition of $\beta$~\eqref{eq:beta_Taylor}. Then the more appropriate way to find the relevant scale for the transition is to compute the bubble size directly from the bubble number density~\cite{Ellis:2018mja}    
\begin{equation}\label{eq:bubble_separation_RD}
n_B = (R_{*\rm R})^{-3}=\int^{t}_{t_c} dt'\, \frac{a(t')^3}{a(t)^3} \, \Gamma(t') P(t') \, .
\end{equation}
Then we indeed find that points with very low $\beta/H$ values always predict large but allowed bubble sizes $H\, R_*<1$, contrary to what one would find naively assuming exponential nucleation.

%%%%%%%%%%%%%%%%%%%%%%%%%%%%%%%%%%%%%%%%%%%%%%%%%%%%%%%%%%%%%%%%%%%%%%%%%%%%%%%%%%%%%%
%%%%%%%%%%%%%%%%%%%%%%%%%%%%%%%%%%%%%%%%%%%%%%%%%%%%%%%%%%%%%%%%%%%%%%%%%%%%%%%%%%%%%%
%%%%%%%%%%%%%%%%%%%%%%%%%%%%%%%%%%%%%%%%%%%%%%%%%%%%%%%%%%%%%%%%%%%%%%%%%%%%%%%%%%%%%%
\section{Analytic phase transition parameters}\label{app:PTparameters}
We quote here the explicit analytical formulae for parameters of the phase transitions in models discussed in Section~\ref{sec:genral_polynomial}. 
\subsection{Polynomial with only a thermal barrier}
We start with the polynomial potential with purely thermal barrier we discussed in Section~\ref{sec:thermal_polynomial}. The strength and duration of the transition read 
\be \label{eq:alpha_TB}
\alpha=\frac{15 \left(\sqrt{9-4 \delta }+3\right) \left(-4 \delta +3 \sqrt{9-4 \delta }+9\right) g_m^2 \left(48 \pi ^2 g_{m^2} \lambda - \delta\,  g_m^2 \right)}{32\times 24^4 \pi
   ^6 \sqrt{9-4 \delta } g_* \lambda ^3}
\ee 
 and
 \be \label{eq:beta_TB}
 \begin{split}
\frac{\beta}{H}  =\frac{2 \sqrt{\delta } \left[-\beta_1 (\delta +6) +\beta_2 \delta (\delta -10)+\beta_3 \delta^2  (3 \delta -14)\right] \left(48 \pi ^2 g_{m^2} \lambda - \delta \, g_m^2 \right)}{243 (\delta -2)^3 g_m \lambda ^{3/2}} \, .
\end{split}
 \ee
These are both independent of temperature, other than through the dimensionless parameter $\delta$.
We an also easily calculate the nucleation temperature, which may be approximated by solving
\be \label{eq:GammaoverH4_TB}
\frac{\Gamma}{H^4}=\frac{T^4 e^{-\frac{S_3}{T}}}{\rho_R+\rho_V}
=
\frac{e^{-\frac{2 g_m \left(\frac{\delta }{\lambda }\right)^{3/2} (\beta_1+\beta_2 \delta+\beta_3 \delta^2 )}{243 (\delta -2)^2}}}
{\left(\frac{T_0^2}{1-\frac{g_m^2 \delta }{48\pi^2 g_{m^2} \lambda }}\right)^2
\left(\frac{1}{3 M_p^2}\right)^2
\left(\frac{\pi ^2 g_*}{30}+
g_m^4\frac{\left(\sqrt{9-4 \delta }+3\right)^2 \left( \sqrt{9-4 \delta }+3-2 \delta\right)  }{128\times 24^4 \pi ^4 \lambda ^3}
\right)^2}
   =1 \, ,
\ee
where we converted the temperature in the Hubble rate into $\delta$ using Eqs.~\eqref{eq:renormalisableV_total},~\eqref{eq:approx_generic_poly_action} and~\eqref{eq:thermalbarrierV}.
This simplifies the calculation significantly, allowing us easily to solve numerically the equation, using the fact that varying the temperature in the interesting range from $T_0$ to $T_c$ simply means varying $\delta$ from $0$ to $2$.

%%%%%%%%%%%%%%%%%%%%%%%%%%%%%%%%%%%%%%%%%%%%%%%%%%%%%%%%%%%%%%%%%%%%%%%%%%%%%%%%%%%%%%%%%%%%%%%%
\subsection{Zero-temperature barrier}\label{app:zeroTbarrier}
We now proceed to the polynomial potential model with a barrier also present at $T=0$, discussed in Section~\ref{sec:treelevel_polynomial}.
The strength and duration of the transition read
\be
 \begin{split}
\alpha & = -\frac{5 \left(\sqrt{9-4 \delta }+3\right) g_{m^2}^2}{512 \pi ^2 \sqrt{9-4 \delta } g_* \lambda  \left(3 A^2 \delta +\lambda  T_0^2 g_{m^2}\right)^2}
\left[3 A^4 \left\{2 \left(\sqrt{9-4 \delta }+6\right) \delta -9 \left(\sqrt{9-4 \delta }+3\right)\right\}+ \right. \\
%\\
 & \qquad \qquad \qquad + \left. A^2 \left\{4 \delta -3 \left(\sqrt{9-4 \delta }+3\right)\right\} \lambda  T_0^2
   g_{m^2}\right]
   \end{split}
\ee 
 and
 \be
 \begin{split}\label{eq:betaoverHZeroTBarrier}
\frac{\beta}{H} & =\frac{8 \pi  \sqrt{\frac{\delta }{\lambda }} g_{m^2} \sqrt{\frac{3 A^2 \delta }{\lambda  g_{m^2}}+T_0^2}
}{243 (\delta -2)^3 \left(3 A^3 \delta +A \lambda  T_0^2 g_{m^2}\right)}
\left[-6 A^2 \delta  \left\{\beta_1 (\delta +2)+  4 \beta_2\delta-\beta_3 \delta^2(\delta -6)  \right\}+ \right. \\
 & \qquad \qquad \qquad - \left. \lambda  T_0^2 g_{m^2}
   \left\{\beta_1 (\delta +6)-\beta_2\delta (\delta -10)+\beta_3\delta^2 (14-3 \delta )\right\}\right] \, .
\end{split}
 \ee
We again express all temperature dependence through the dimensionless parameter $\delta$.
To calculate the nucleation temperature of the bubbles we need to solve
\be \label{eq:GammaoverH4_zeroTbarrier}
\begin{split}
\frac{\Gamma}{H^4} & =\frac{T^4 e^{-\frac{S_3}{T}}}{\rho_R+\rho_V}
= \\
& =\frac{
\left(\frac{3 A^2 \delta }{\lambda  g_{m^2}}+T_0^2\right)^2
e^{
\frac{8 \pi  A \left(\frac{\delta }{\lambda }\right)^{3/2} (\text{$\beta $1}+\delta  (\text{$\beta $2}+\text{$\beta $3} \delta ))}{81 (\delta -2)^2 \sqrt{\frac{3 A^2 \delta }{\lambda  g_{m^2}}+T_0^2}}
}}
{
\left(\frac{1}{3 M_p^2}\right)^2
\left(\frac{\pi ^2}{30}  g_* \left(\frac{3 A^2 \delta }{\lambda  g_{m^2}}+T_0^2\right)^2+\frac{A^4 \left(\sqrt{\frac{4 \lambda  T_0^2 g_{m^2}}{3 A^2}+9}+3\right){}^2 \left(\frac{2 \lambda  T_0^2 g_{m^2}}{3 A^2}+\sqrt{\frac{4 \lambda  T_0^2
   g_{m^2}}{3 A^2}+9}+3\right)}{2048 \lambda ^3}\right)^2
}
=1 \, ,
\end{split}
\ee
where we converted the temperature in the Hubble rate into $\delta$ using Eqs.~\eqref{eq:renormalisableV_total},~\eqref{eq:approx_generic_poly_action} and~\eqref{eq:zeroTbarrierV}.
The calculation is slightly more complicated due to presence of a barrier at tree level. Essentially, one needs to find a minimum of Eq.~\eqref{eq:GammaoverH4_zeroTbarrier} setting the minimal value of $\delta_{\rm min}$, which might not exist if the barrier is too large and tunnelling too suppressed. If this is not the case and $\Gamma/H|_{\delta=\delta_{\rm min}}<1$, we can find a solution varying $\delta$ from $\delta_{\rm min}$ to $2$.

%%%%%%%%%%%%%%%%%%%%%%%%%%%%%%%%%%%%%%%%%%%%%%%%%%%%%%%%%%%%%%%%%%%%%%%%%%%%%%%%%%%
\subsection{Classically scale-invariant potential}\label{app:scaleinvariant}
Finally, we proceed to the case of classically scale-invariant potential discussed in Section~\ref{sec:scaleinvariant}.
We can express the key parameters of the transition as
\be
\begin{split}
\alpha & =\frac{5(1-\delta)}{g_* \delta^2}\, , \\
\frac{\beta}{H} & =
\frac{\pi ^3 I^3 2^{n_{\mu }+6} \delta ^{n_{\mu }-\frac{1}{2}}}{9 g^3 v (2 \delta -1)^3 } \left[2 \delta ^4 \mu _3-\delta ^3 \left(2 \mu _2+5 \mu _3\right)-3 \delta ^2 \left(2 \mu _1+\mu _2\right)+ \right. \\
 & \left.-\delta  \left(\mu _1+10\right) +2 (2 \delta -1) \left(\delta ^3 \mu _3+\delta ^2
   \mu _2+\delta  \mu _1+1\right) n_{\mu }+1\right] \, .
   \end{split}
\ee
We can also determine the nucleation temperature by solving
\be \label{eq:GammaoverH4_ScaleInv}
\begin{split}
\frac{\Gamma}{H^4} =\frac{T^4 e^{-\frac{S_3}{T}}}{\rho_R+\rho_V}
 =\frac{
\frac{9 \delta ^2 g^4 v^4}{4 \pi ^4}
e^{-
\frac{64\pi ^3 I^3 (2\delta) ^{n_{\mu }}  \left(1+\mu _1 \delta +\mu _2 \delta^2+\mu _3 \delta^3\right) }{9  g^3 v \sqrt{\delta}(1-2 \delta )^2}
}}
{
\left(\frac{1}{3 M_p^2}\right)^2
\left(
\frac{\pi ^2}{30}  g_* \frac{9 \delta ^2 g^4 v^4}{4 \pi ^4}
+
\frac{3 (1-2 \delta ) g^4 v^4}{8 \pi ^2}
   \right)^2
}
=1 \, ,
\end{split}
\ee
where we again converted the temperature into $\delta$ using Eq.~\eqref{eq:SIpotential}. As in the previous cases, we have to solve the above numerically, varying from $\delta=0$, which corresponds to zero temperature, to  $\delta=1/2$, which corresponds to the critical temperature.

\bibliographystyle{JHEP}
\bibliography{SoundWaves}
\end{document}